\documentclass[times,twocolumn]{aastex631}
\usepackage{graphicx}
\usepackage[flushleft]{threeparttable}
\usepackage{blindtext}
\usepackage{amsmath}
\usepackage{mathtools}
\usepackage{multirow}
\usepackage{comment}
\turnoffeditone

\DeclareRobustCommand{\ion}[2]{%
\relax\ifmmode
\ifx\testbx\f@series
{\mathbf{#1\,\mathsc{#2}}}\else
{\mathrm{#1\,\mathsc{#2}}}\fi
\else\textup{#1\,{\mdseries\textsc{#2}}}%
\fi}

\begin{document}
\shortauthors{Jung et al.}
\def\nar{New Astron.}
\def\na{New Astron.}
\title{\large \textbf{Constraints on the Lyman Continuum Escape from Low-mass Lensed Galaxies at 1.3 $\leq$ \textit{z} $\leq$ 3.0}}

\correspondingauthor{Intae Jung}
\email{ijung@stsci.edu}

\author[0000-0003-1187-4240]{Intae Jung}
\affil{Space Telescope Science Institute, 3700 San Martin Drive Baltimore, MD 21218, United States}

\author[0000-0001-7113-2738]{Henry C. Ferguson}
\affil{Space Telescope Science Institute, 3700 San Martin Drive Baltimore, MD 21218, United States}

\author[0000-0001-8587-218X]{Matthew J. Hayes}
\affiliation{Stockholm University, Department of Astronomy and Oskar Klein Centre for Cosmoparticle Physics, AlbaNova University Centre, SE-10691, Stockholm, Sweden.}

\author[0000-0002-6586-4446]{Alaina Henry}
\affiliation{Center for Astrophysical Sciences, Department of Physics \& Astronomy, Johns Hopkins University, Baltimore, MD 21218, USA}
\affil{Space Telescope Science Institute, 3700 San Martin Drive Baltimore, MD 21218, United States}

\author[0000-0002-6790-5125]{Anne E. Jaskot}
\affiliation{Department of Astronomy, Williams College, Williamstown, MA 01267, USA}

\author[0000-0001-7144-7182]{Daniel Schaerer}
\affiliation{Observatoire de Gen\`eve, Universit\'e de Gen\`eve, Chemin Pegasi 51, 1290 Versoix, Switzerland}
\affiliation{CNRS, IRAP, 14 Avenue E. Belin, 31400 Toulouse, France}

\author[0000-0002-7559-0864]{Keren Sharon}
\affiliation{Department of Astronomy, University of Michigan, 1085 S. University Ave, Ann Arbor, MI 48109, USA}

\author[0000-0001-5758-1000]{Ricardo O. Amor\'{i}n}
\affiliation{ARAID Foundation. Centro de Estudios de F\'{\i}sica del Cosmos de Arag\'{o}n (CEFCA), Unidad Asociada al CSIC, Plaza San Juan 1, E--44001 Teruel, Spain}
\affiliation{Departamento de Astronom\'{i}a, Universidad de La Serena, Av. Juan Cisternas 1200 Norte, La Serena 1720236, Chile}

\author[0000-0002-7570-0824]{Hakim Atek}
\affiliation{Institut d'Astrophysique de Paris, CNRS, Sorbonne Universit\'e, 98bis Boulevard Arago, 75014, Paris, France}

\author[0000-0003-1074-4807]{Matthew B. Bayliss}
\affiliation{Department of Physics, University of Cincinnati, Cincinnati, OH 45221, USA}

\author[0000-0003-2200-5606]{H\r{a}kon Dahle} 
\affiliation{Institute of Theoretical Astrophysics, University of Oslo, P.O. Box 1029, Blindern, NO-0315 Oslo, Norway}

\author[0000-0001-8519-1130]{Steven L. Finkelstein}
\affiliation{Department of Astronomy, The University of Texas at Austin, Austin, TX, USA}

\author[0000-0002-5688-0663]{Andrea Grazian}
\affil{INAF--Osservatorio Astronomico di Padova,Vicolo dell'Osservatorio 5, I-35122, Padova, Italy}

\author[0000-0002-4902-0075]{Lucia Guaita}
\affil{Universidad Andres Bello, Facultad de Ciencias Exactas, Departamento de Fisica, Instituto de Astrofisica, Fernandez Concha 700, Las Condes, Santiago RM, Chile}

\author[0000-0002-3005-1349]{Göran Östlin}
\affiliation{Department of Astronomy and Oskar Klein Centre for Cosmoparticle Physics, AlbaNova University Centre, Stockholm University, SE-10691, Stockholm, Sweden}

\author[0000-0001-8940-6768]{Laura Pentericci}
\affiliation{INAF - Osservatorio Astronomico di Roma, via di Frascati 33, 00078 Monte Porzio Catone, Italy}

\author[0000-0002-5269-6527]{Swara Ravindranath}
\affiliation{Astrophysics Science Division, NASA Goddard Space Flight Center, 8800 Greenbelt Road, Greenbelt, MD 20771, USA}
\affiliation{Center for Research and Exploration in Space Science and Technology II, Department of Physics, Catholic University of America, 620 Michigan Ave N.E., Washington DC 20064, USA}

\author[0000-0002-9136-8876]{Claudia Scarlata}
\affil{Minnesota Institute for Astrophysics, University of Minnesota, 316 Church st. SE Minneapolis, MN 55455, United States} 

\author[0000-0002-7064-5424]{Harry I. Teplitz}
\affiliation{IPAC, Mail Code 314-6, California Institute of Technology, 1200 E. California Blvd., Pasadena CA, 91125, USA}

\author[0000-0002-2201-1865]{Anne Verhamme}
\affiliation{Observatoire de Gen\`eve, Universit\'e de Gen\`eve, Chemin Pegasi 51, 1290 Versoix, Switzerland}

\submitjournal{the Astrophysical Journal}

\begin{abstract}
Low-mass galaxies can significantly contribute to reionization due to their potentially high Lyman continuum (LyC) escape fraction and relatively high space density.
We present a constraint on the LyC escape fraction from low-mass galaxies at $z=1.3$ -- $3.0$.
We obtained rest-frame UV continuum imaging with the ACS/SBC and the WFC3/UVIS from the Hubble Space Telescope for eight strongly-lensed galaxies that were identified in the Sloan Giant Arc Survey (SGAS) and the Cluster Lensing And Supernova survey with Hubble (CLASH). 
The targeted galaxies were selected to be spectroscopically confirmed, highly magnified, and blue in their UV spectral shapes ($\beta<-1.7$). Our targets include intrinsically low luminosity galaxies down to a magnification-corrected absolute UV magnitude of $M_{\rm UV}\sim-14$.
We perform custom-defined aperture photometry to place the most reliable upper limits of LyC escape from our sample. 
From our observations, we report no significant ($>$\,$2\sigma$) detections of LyC fluxes, placing 1$\sigma$ upper limits on the absolute LyC escape fractions of 3 -- 15\%.
Our observations do not support the expected increased escape fractions of LyC photons from intrinsically UV faint sources. 
Considering the highly anisotropic geometry of LyC escape, increasing the sample size of faint galaxies in future LyC observations is crucial.
\end{abstract}
\keywords{Reionization (1383); Extragalactic astronomy (506); Dwarf galaxies (416); Ultraviolet astronomy (1736); Strong gravitational lensing(1643)}

\section{Introduction}
Over the redshift range of around 15 to 6, most of the hydrogen gas in the intergalactic medium (IGM) changed from neutral to ionized. The conventional thought is that young stars in newly-formed galaxies played a crucial role in emitting ionizing photons, although a smaller portion likely came from accreting black holes \citep[or active galactic nuclei; AGNs, see the review of ][and the references therein]{Robertson2022a}. A flood of information provided by JWST observations is shedding light on galaxies and AGNs during this era of reionization, including direct spectroscopic confirmation of candidate galaxies as well as discovery of $z\gtrsim9$ AGNs \citep[e.g.,][]{Roberts-Borsani2022a, Curtis-Lake2022a, Fujimoto2023a, Bunker2023a, Arrabal-Haro2023a, Arrabal-Haro2023b, Harikane2023a, Larson2023a, Maiolino2023a}.

Measurements of the galaxy ultraviolet (UV) luminosity function at $z\gtrsim7$ constrain the number densities of the likely sources of ionizing photons in relation to their absolute UV magnitude \citep[e.g.,][]{Donnan2023a, Harikane2023a, Finkelstein2023b, McLeod2023a}.  The spectral-energy distributions (SEDs) of galaxies can be used with population-synthesis models to indirectly infer the production rate of ionizing photons, often parameterized by $\xi_{\rm ion}$, the ionizing-photon production rate relative to 1500\AA\ luminosity $L_{\nu}$ \citep[e.g.,][]{Madau1999a, Schaerer2003a, Bouwens2012a}.  Whereas emission line information allows more direct measurements of $\xi_{\rm ion}$, comparing the line fluxes to the observed UV \citep[e.g.,][]{Bouwens2016a, Nakajima2016a, Matthee2017a, Shivaei2018a, Lam2019a}.  With access to rest-optical wavelength spectra of reionization-era galaxies given by JWST observations, now it also becomes possible to derive $\xi_{\rm ion}$ from the relative strengths of Balmer emission lines to the 1500\AA\ luminosity \cite[e.g.,][]{Jung2023a, Atek2023a}. 

Nonetheless, it is challenging to precisely determine the escape of these crucial photons and their contribution to reionization. To accurately assess ionizing photon supplies from galaxies during the reionization era, a comprehensive understanding of the escape fraction for ionizing (or Lyman-continuum; LyC) photons originating from these galaxies is essential.  Unfortunately, as we venture into the epoch of reionization, the direct measurement of LyC escape becomes infeasible due to the attenuation imposed by the IGM \citep{Madau1995a, Inoue2014a}. 

The best alternative is to use lower-redshift star-forming galaxies.  The systematic trends observed in samples of these low-redshift galaxies can help reveal the conditions that are conducive to LyC escape. Low-redshift samples can also be used to calibrate a set of proxy measurements that might reliably predict the emergent ionizing continuum from a galaxy. UV observations with the Hubble Space Telescope (HST) of $z\sim0.3$ compact galaxies that are considered as reionization-era analogs have successfully detected LyC emission \citep[e.g.,][]{Izotov2016a, Izotov2018a, Wang2019a, Flury2022a}. Recently, the Low-redshift Lyman-Continuum Survey (LzLCS; PI: A. Jaskot) has significantly enlarged the sample of LyC emitters \citep{Flury2022a}.  

These LyC observations enable the investigation of correlations between LyC escape and other galaxy properties that can serve as indirect indicators of LyC escape. This includes, for example, the escape of Ly$\alpha$, the [\ion{O}{iii}]/[\ion{O}{ii}] line ratio, \ion{Mg}{ii} emission, rest-UV $\beta$ slope, and galaxy sizes \citep[e.g.,][]{Dijkstra2016a, Verhamme2015a, Henry2018a, Izotov2018a, Izotov2020a, Chisholm2018a, Chisholm2020a, Chisholm2022a, Saldana-Lopez2022a, Xu2022a, Xu2023a}. The recent increase in the size and diversity of low-redshift samples has made it possible to undertake a multivariate analysis that substantially reduces the scatter in predicted escape fractions relative to predictions based on a single observable \citep[][and Jaskot et al. in preparation]{Flury2022b, Mascia2023a}. JWST observations are beginning to provide access to some of these indirect indicators of LyC escape fraction ($f_{\rm esc}$) for reionization-era galaxies \citep{Cameron2023a, Saxena2023b}. The data will get much more extensive as spectroscopic data accumulate. There is thus a realistic prospect of applying these indirect indicators to make a more accurate estimate of the contribution of individual galaxies to the ionizing photon budget \citep{Mascia2023a, Mascia2023b, Jung2023a}. However, the current $f_{\rm esc}$ predictions rely on the correlations between indirect indicators and $f_{\rm esc}$ that show generally large scatter and need further testing. This requires an increased dataset of direct LyC observations that can be measured at lower redshifts. 

Specifically, the applicability of $z\sim0.3$ relationships to the reionization epoch remains uncertain. Therefore, conducting this study at a modest redshift is paramount to enhancing our understanding, and there has been recent progress on identifying LyC emitters at higher redshift at $z>2$ \citep{Mostardi2015a, deBarros2016a, Shapley2016a, Vanzella2016a, Bian2017a, Vanzella2018a, Steidel2018a, Fletcher2019a, Ji2020a}. However, many of these are high luminosity galaxies with $M_{\rm UV}\lesssim-19$, comparable to the limit in LzLCS. The LyC contribution may be more significant from lower luminosity galaxies, which have hotter stellar populations and may be more capable of clearing neutral gas, at least along some sightlines \citep[e.g.,][]{Wise2014a, Paardekooper2015a, Xu2016a, Anderson2017a}. To fill in the parameter space of calibrating galaxies, the most important complement to the existing data is to measure lower-luminosity galaxies at $z \gg 0$. 

The most important redshift range is $1 < z < 3.5$, where the IGM opacity is low enough to allow us to detect LyC radiation and measure the evolutionary trends. This has been attempted with mixed success.  There have been successes with detecting leaking LyC photons at $z = 2$ -- $3$ but mostly from luminous sources \citep[e.g.,][]{Steidel2001a, Steidel2018a, Shapley2006a, Iwata2009a, Shapley2016a, Vanzella2012a, Vanzella2016a, Vanzella2018a, Mostardi2015a, deBarros2016a, Bian2017a, Saha2020a, Pahl2021a, Marques-Chaves2021a, Begley2022a}. Meanwhile, many other deep observations found much lower LyC escape fractions, mostly delivering upper limits \citep[e.g.,][]{Siana2010a, Rutkowski2016a, Rutkowski2017a, Grazian2017a, Fletcher2019a, Alavi2020a}. A more comprehensive overview of previous efforts on detecting LyC leakage at this redshift is available in \cite{Kerutt2023a}.

Focusing on the observations of lensed galaxies, the Sunburst Arc, a lensed dwarf galaxy at $z = 2.37$, presents an averaged LyC escape fraction of $f_{\rm esc,abs} = 32$\% \citep{Rivera-Thorsen2017a, Rivera-Thorsen2019a, Vanzella2020a, Mainali2022a, Kim2023a}. However, \cite{Amorin2014a} targeted a strongly lensed sub-$L^*$ extreme emission line galaxy at $z = 3.417$, but finding no clue of $f_{\rm esc,rel}\gtrsim20$\%. \cite{Vasei2016a} obtained $f_{\rm esc,rel}\lesssim$ 8\% (corresponding to $f_{\rm esc,abs}\lesssim$ 2\%) for the Cosmic Horseshoe (magnification $\mu\sim24$). \cite{Bian2017a} estimate $f_{\rm esc} = 28$ - 57\% from a lensed arc at $z = 2.5$ in A2218, but this was the only one of the seven highly magnified arcs they observed that was detected. The six Frontier Fields and cluster A1689 were observed for 6-8 orbits with WFC3/F275W (Program ID 14209, PI: B. Siana) to detect escaping LyC continuum at $z\sim2.4$. There have been no reports of detections or upper limits, although the analysis is limited only to a few spectroscopically confirmed galaxies. More LyC-leaking galaxies may be found as redshift surveys improve.

In this paper, we present our HST observations aiming to detect leaking LyC photons from eight galaxies strongly boosted by gravitational lensing. We selected highly magnified spectroscopically-confirmed sources with blue rest-UV slopes and source plane absolute magnitudes $M_{\rm UV} < -19$, based on the lens models available when the observations were proposed. We constrain LyC escape fractions of our eight sources at $1.3\leq z \leq3.0$ including intrinsically low luminosity galaxies down to a magnification-corrected $M_{\rm UV}\sim-14$.

This paper is structured as follows. We describe the HST imaging data, data reduction, and photometry of our targets in Section 2. Section 3 presents measurements of LyC escape as well as galaxy physical properties. We discuss our results in Section 4, compared to other studies in the literature. We summarize our findings and conclude in Section 5. 

Throughout this work, we assume the Planck cosmology \citep{Planck-Collaboration2016a} with $H_0$ = 67.8\,km\,s$^{-1}$\,Mpc$^{-1}$, $\Omega_{\text{M}}$ = 0.308, and $\Omega_{\Lambda}$ = 0.692. All flux densities are given in $f_{\nu}$, and all magnitudes are quoted in the AB system \citep{Oke1983a}.

\begin{deluxetable*}{ccccccccc}
\label{tab:targets}
\tabletypesize{\footnotesize}
\tablecaption{Lensed Galaxy Sample with the LyC Observation Summary} 
\tablehead{
\colhead{ID} & \colhead{R.A. (J2000.0)} & \colhead{Decl. (J2000.0)}  & \colhead{$z_{\rm cluster}$} & \colhead{$z_{\rm galaxy}$} & \colhead{$\mu$} & {LyC Filter} & {$t_{\rm exp}$ (hr)} & {Observation Dates} \\
\colhead{(1)} & \colhead{(2)} & \colhead{(3)} & \colhead{(4)} & \colhead{(5)} & \colhead{(6)} & \colhead{(7)} & \colhead{(8)} & \colhead{(9)}
}
\startdata
{SDSS J0851+3331-B} & {132.9082792} & {33.51857778} & {0.3689} & {1.3454} & {$63.9^{+5.4}_{-28.2}$} & {SBC F150LP} & {3.0} & {2018 Oct 31, 2019 Mar 16} \\
{SDSS J0915+3826-A} & {138.9089458} & {38.45128333} & {0.3961}  & {1.501} & {$81.4^{+25.3}_{-23.4}$} & {SBC F150LP} & {2.25} & {2018 Nov 21, 2019 May 7} \\
{SDSS J1138+2754-A} & {174.5363208} & {27.91240278} & {0.451}  & {1.3335} & {$6^{+0}_{-0}$} & {SBC F150LP} & {4.9} & {2019 Mar 26, Apr 5,6,8} \\
{SDSS J1138+2754-C} & {174.5367917} & {27.91420278} & {0.451}  & {1.455} & {$198.3^{+187.7}_{-144.1}$} & {SBC F150LP} & {4.9} & {2019 Mar 26, Apr 5,6,8}  \\
{SDSS J1439+1208-A} & {219.7926625} & {12.13835833} & {0.42734}  & {1.494} & {$456.3^{+78.4}_{-155.0}$} & {SBC F150LP} & {2.25}& {2018 Sep 4} \\
{M0329-2556} & {52.429404} & {-2.194035} & {0.450}  & {1.313} & {$5.4^{+0.3}_{-0.4}$}  & {SBC F150LP} & {4.5} & {2019 Sep 16, 19}  \\
{M1115-2511} & {168.955867} & {1.497984} & {0.352}  & {1.599} & {$8.0^{+0.7}_{-0.4}$}   & {SBC F150LP} & {6.0}& {2019 May 17,18,19} \\
{M1206-4340} & {181.560201} & {-8.809416} & {0.440}  & {3.036} & {$5.1^{+0.2}_{-0.3}$}  & {UVIS F336W} & {6.0}& {2019 May 21,25,26,29} \\
\enddata
\tablecomments{Col.(1) Object ID, (2) Right ascension, (3) Declination, (4) Cluster redshifts \citep{Postman2012a, Sharon2020a}, (5) Galaxy spectroscopic redshifts \citep{Bayliss2011a, Sharon2020a, Caminha2017a, Caminha2019a}, (6) Gravitational lensing magnification, (7) HST imaging filter used to observe LyC fluxes, (8) Total exposure time, (9) HST observation dates}
\end{deluxetable*}

\section{Data and Photometry}
\subsection{Targets}
We targeted eight galaxies that are highly magnified by gravitational lensing. All eight galaxies have spectroscopically measured redshifts and blue rest-UV colors. The targets were identified in the Sloan Giant Arc Survey (SGAS; PI: Gladders; \citealt{Bayliss2011a, Sharon2020a}) and the Cluster Lensing And Supernova survey with Hubble (CLASH; PI: M. Postman), including intrinsically lower luminosity galaxies than previous studies. The targets mostly were detected in their strong emission lines (e.g., [\ion{O}{ii}]$\lambda\lambda$3726,3729) from ground-based observations \citep{Bayliss2011a, Sharon2020a, Caminha2019a} except for M1206-4340.  Strong emission lines are deficient in the MUSE spectra of M1206-4340. Instead, M1206-4340 presents strong absorption features with extended Ly$\alpha$ blobs around the source \citep{Zitrin2012a, Caminha2017a}.  The sources span redshifts of $1.3\leq z\leq3.0$, and their principal properties are listed in Table \ref{tab:targets}.

\subsection{HST Observations}
The primary goal of this work is to detect LyC emission from intrinsically less luminous but highly magnified galaxies through gravitational lensing.  We obtained rest-frame LyC ($\lambda<912$\AA) imaging using the F150LP filter of SBC on ACS for seven $z\simeq1.3$\,--\,$1.6$ targets and the F336W filter of UVIS on WFC3 for a $z=3$ target (Program ID 15271, PI: H. Ferguson). The F150LP filter has a significant throughput ($>$1\%) at a wavelength range of $1464$\AA\ $<\lambda<1734$\AA\ with a pivot wavelength ($\lambda_{\rm pivot}$) at 1605.7\AA, which corresponds to $\lambda_{\rm rest}\sim620$\,--\,700\AA\ at $z\simeq1.3$\,--\,$1.6$ \citep{Siana2010a, Alavi2020a}. The F336W filter covers $\lambda_{\rm rest}\sim838$\AA\ at $z=3.0$, having $\lambda_{\rm pivot}=3355$\AA\ with a $>$1\% throughput at $3016$\AA\ $<\lambda<3708$ \AA.
The integration time varies depending on the observed apparent magnitude in a band corresponding to the rest-frame 1500\AA\, ensuring reaching the depth to detect LyC photons with a desired detection significance.

We also use the archival HST multi-band imaging data for our sources. Typically, four HST-broadband-filter observations are available for SGAS sources from the HST program GO-13003 (PI: Gladders), and the CLASH targets have sixteen broadband-filter-imaging data. We downloaded the fully reduced science mosaic images from the MAST archive for SGAS\footnote{https://archive.stsci.edu/hlsp/sgas} and CLASH\footnote{https://archive.stsci.edu/prepds/clash/}. For details of filter selection, observational strategy, and data reduction, we refer the reader to \cite{Sharon2020a} and \cite{Postman2012a} for SGAS and CLASH, respectively. 

\subsection{Data Reduction of LyC Observations}
We first downloaded the drizzled images (\texttt{drz.fits} files) from the Mikulski Archive for Space Telescopes (MAST) portal for F150LP filter observations (Program ID 15271, PI: H. Ferguson). The images were registered to the existing optical HST images using individually measured positions of sources in the UV images. The reference positions in the world-coordinate systems (WCS) for the UV images were updated to account for the relatively small shifts due to the use of different guide stars in the different HST observations. The registered individual images were then combined using the drizzle algorithm \citep{Fruchter2002a} to produce a single co-added image per target. 

It is known that ACS/SBC data suffer time-dependent dark current rates, which often cause spatial structures in a background \citep{Alavi2020a, Runnholm2023a}. To mitigate the effect, we subtracted background substructures from the downloaded drizzled images by estimating a background from bright-sources-masked images using the \texttt{Background2D} class in the {\sc background} subpackage in {\sc PHOTUTILS} \citep{Bradley2023a}.

For F336W observations, we downloaded the flat-field CTE-corrected images (\texttt{flc.fits} files) to perform better cosmic-ray (CR) removal as the reduced drizzled images of F336W available from the MAST contain numerous CRs uncleaned. We used the {\sc Astrodrizzle} task within the {\sc Drizzlepac} package\footnote{https://drizzlepac.readthedocs.io/en/latest/index.html} to combine individual exposures and create the drizzled image (\texttt{drc.fits}). The {\sc Astrodrizzle} task applies the geometric distortion correction and masks bad pixels that are typically CRs or faulty pixels in the detector. In {\sc Astrodrizzle}, first, undistorted images are generated using the astrometric and distortion information in the header of \texttt{flc.fits} files. Then, CRs are identified in individual \texttt{flc.fits} frames by comparing them to the median image. We adjusted the \textit{driz\_cr\_snr} parameters of the CR-detection thresholds and opted for \textit{driz\_cr\_snr} $=$ [5,1.5]. The first threshold in \textit{driz\_cr\_snr} was chosen to avoid too many false positives but finding obvious CRs, and the second threshold was selected low enough to capture the faint peripherals of CRs \citep{Zhang2020a}. We created the drizzled image with the same natural pixel scale (0.04\arcsec) of the input \texttt{flc} images.

\begin{figure*}[th]
\centering
\includegraphics[width=\textwidth]{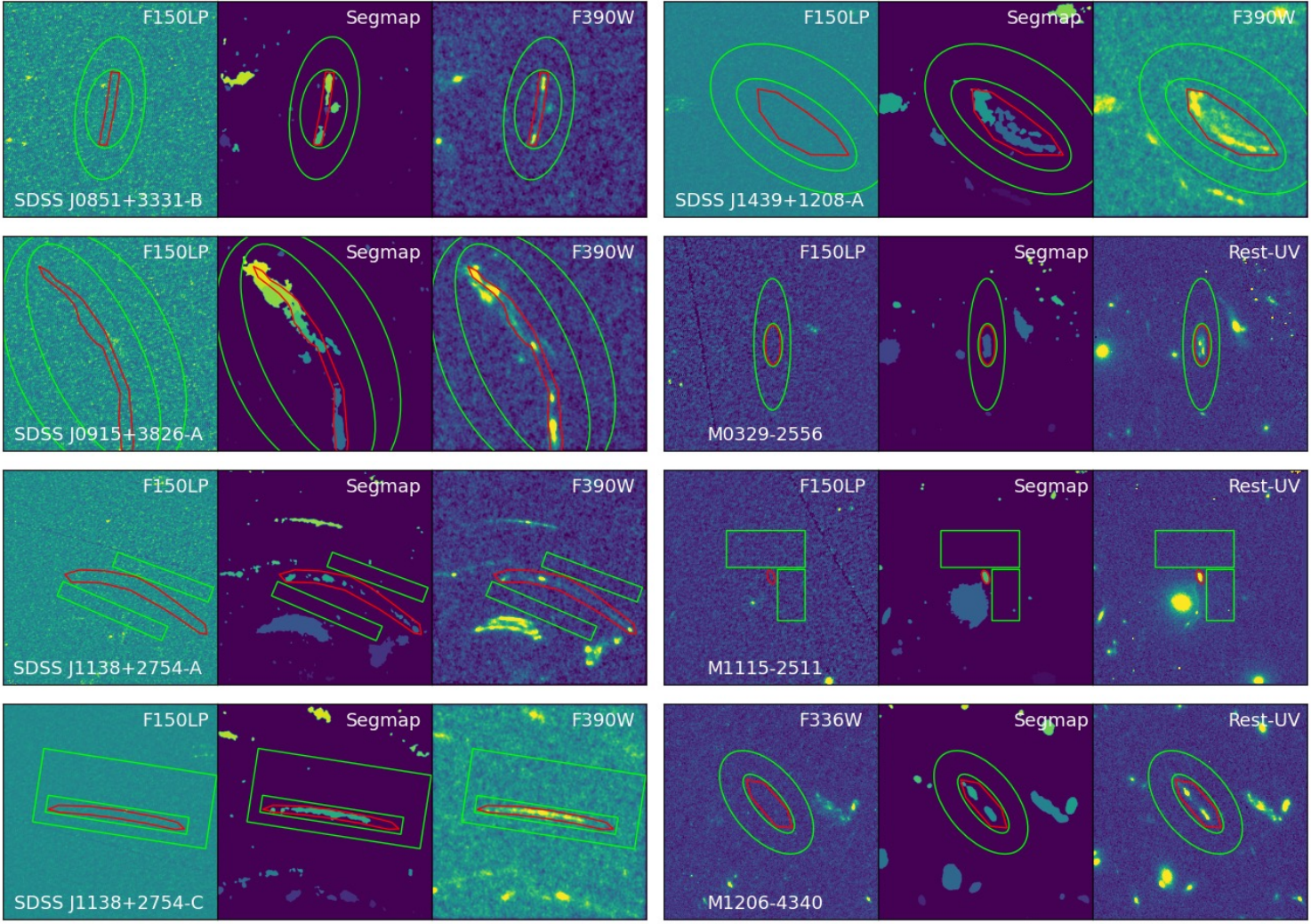}
\caption{$17\arcsec\times17\arcsec$ HST Postages of our eight targets. In each target, from left to right, we present LyC filter imaging, segmentation maps used for LyC photometry, and a HST broadband filter covering the rest-UV of the targets. In the panels, the red polygons show the tailored boundaries for our broadband photometry, and the background regions are marked as green.} 
\label{fig:HST_cutouts}
\end{figure*}

\begin{deluxetable*}{ccccccc}
\label{tab:sgas_targets}
\tabletypesize{\footnotesize}
\tablecaption{HST Customized Aperture Photometry of the SGAS targets} 
\tablehead{
\colhead{ID} & \colhead{WFC3/UVIS F390W} & \colhead{WFC3/UVIS F775W} & \colhead{WFC3/UVIS F814W} &  \colhead{WFC3/IR F110W} & \colhead{WFC3/IR F125W} & \colhead{WFC3/IR F160W}
}
\startdata
{SDSS J0851+3331-B} & {$1.34 \pm 0.30$} & {-} & {$1.84 \pm 0.35$} & {-} & {$3.44 \pm 0.11$} & {$2.92 \pm 0.14$} \\
{SDSS J0915+3826-A} & {$8.84 \pm 0.56$} & {-} & {$11.87 \pm 0.65$} & {-} & {$22.85 \pm 0.21$} & {$23.84 \pm 0.26$} \\
{SDSS J1138+2754-A} & {$4.07 \pm 0.78$} & {$6.04 \pm 1.15$} & {-} & {$11.84 \pm 0.17$} & {-} & {$14.50 \pm 0.37$} \\
{SDSS J1138+2754-C} & {$1.23 \pm 0.39$} & {$1.01 \pm 0.57$} & {-} & {$1.57 \pm 0.09$} & {-} & {$1.47 \pm 0.19$} \\
{SDSS J1439+1208-A} & {$6.46 \pm 0.65$} & {-} & {$5.57\pm 0.76$} & {-} & {$10.13 \pm 0.25$} & {$9.76 \pm 0.31$} 
\enddata
\tablecomments{All flux densities are given in $\mu$Jy.}
\end{deluxetable*}

\subsection{Broadband Photometry}
It is important that multiband photometry be done with images that have the same point-spread function (PSF). We generated kernels to match the PSF sizes to that of the longest wavelength image (i.e., F160W image).  The CLASH collaboration provides a set of PSF models\footnote{https://archive.stsci.edu/missions/hlsp/clash/psf\_models/} for all 16 filters used in the CLASH observations from F225W to F160W.   However, currently, no public PSF model is available for F150LP. Thus, we construct the F150LP PSF model separately. We downloaded the latest F150LP calibration-field observations for NGC6681 with $>$2000s exposure time (Program ID: 11387).  We selected eight isolated stars that do not have neighboring stars within $\sim$2.5\arcsec distance. Then, we built the F150LP PSF model using the \texttt{EPSFBuilder} class in the {\sc Photutils.psf} subpackage. Finally, we checked the encircled energy curve within a 2.5\arcsec radius, which agrees well with the encircled energy curve available on the STScI webpage\footnote{https://www.stsci.edu/hst/instrumentation/acs/data-analysis/aperture-corrections}.  We present the constructed PSF model and a comparison of the encircled energy curves in the Appendix.

With the PSF models, we created 2-dimensional (2D) kernels using the \texttt{create\_matching\_kernel} function of the PSF Matching subpackage in {\sc Photutils}.  Then, we confirmed that the convolved PSFs agree well in their encircled energy plots.

Typical source-detection schemes do not generally perform well for our sources as the sources are strongly lensed, showing highly stretched and elongated morphologies with clumpy structures. Thus, photometric fluxes in the PSF-matched images were measured within custom-defined areas, ensuring extended source boundaries were included. As our lensed targets are stretched with significant diffuse light, we draw a boundary encompassing the diffuse components while minimizing contributions from nearby contaminants. These tailored boundaries are shown as red polygons in Figure \ref{fig:HST_cutouts}. For an improved background subtraction, we also define field areas in the close vicinity with no obvious neighboring sources. These background regions are noted as green.

The fluxes in individual broadband filters are measured by summing all the fluxes included in the tailored boundaries, and the uncertainties are propagated within the same boundaries using the weight maps. The photometry is corrected for Galactic extinction using the \textit{E(B-V)} reddening values estimated in \cite{Schlegel1998a}, ranging from 0.022 to 0.032 for SGAS images and 0.039 to 0.062 for CLASH images. We assume the \cite{Pei1992a} dust extinction curve with $R_V=3.08$ to correct the Galactic extinction in each broadband filter. The flux values measured from our customized aperture photometry are listed in Table \ref{tab:sgas_targets} for SGAS, and Table \ref{tab:clash_targets} includes our HST photometry of the three CLASH targets.

\begin{deluxetable}{cccc}
\label{tab:clash_targets}
\tabletypesize{\footnotesize}
\tablecaption{Customized Aperture Photometry of the CLASH targets} 
\tablehead{
\colhead{Filters} & \colhead{M0329-2556} & \colhead{M1115-2511} & \colhead{M1206-4340}
}
\startdata
{WFC3/UVIS F225W} & {$0.57 \pm 0.14$} & {$0.09 \pm 0.10$} & {\nodata}  \\
{WFC3/UVIS F275W} & {$0.65 \pm 0.13^{*}$} & {$0.42 \pm 0.08$} & {\nodata} \\
{WFC3/UVIS F336W} & {$0.88 \pm 0.08$} & {$0.72 \pm 0.05^{*}$} & {\nodata} \\
{WFC3/UVIS F390W} & {$1.12 \pm 0.04$} & {$0.97 \pm 0.03$} & {$0.73 \pm 0.04$} \\
{ACS/WFC F435W} & {$1.19 \pm 0.03$} & {$1.00 \pm 0.02$} & {$1.23 \pm 0.03$} \\
{ACS/WFC F475W} & {$1.15 \pm 0.02$} & {$1.02 \pm 0.01$} & {$1.56 \pm 0.02^{*}$} \\
{ACS/WFC F606W} & {$1.19 \pm 0.01$} & {$1.08 \pm 0.01$} & {$2.48 \pm 0.01^{*}$} \\
{ACS/WFC F625W} & {$1.11 \pm 0.02$} & {$1.05 \pm 0.02$} & {$2.93 \pm 0.02$} \\
{ACS/WFC F775W} & {$1.22 \pm 0.03$} & {$1.08 \pm 0.02$} & {$3.11 \pm 0.03$} \\
{ACS/WFC F814W} & {$1.40 \pm 0.02$} & {$1.04 \pm 0.01$} & {$3.10 \pm 0.02$} \\
{ACS/WFC F850LP}& {$1.81 \pm 0.06$} & {$1.16 \pm 0.04$} & {$3.01 \pm 0.06$} \\
{WFC3/IR F105W} & {$2.09 \pm 0.02$} & {$1.18 \pm 0.01$} & {$2.65 \pm 0.02$} \\
{WFC3/IR F110W} & {$1.97 \pm 0.01$} & {$1.61 \pm 0.01$} & {$2.59 \pm 0.01$} \\
{WFC3/IR F125W} & {$2.03 \pm 0.02$} & {$1.69 \pm 0.01$} & {$2.54 \pm 0.02$} \\
{WFC3/IR F140W} & {$2.10 \pm 0.01$} & {$1.68 \pm 0.01$} & {$2.77 \pm 0.01$} \\
{WFC3/IR F160W} & {$2.30 \pm 0.02$} & {$1.22 \pm 0.01$} & {$3.02 \pm 0.02$} 
\enddata
\tablecomments{All flux densities are given in $\mu$Jy.}
\tablenotetext{}{
$^{*}$\footnotesize Filters including Ly$\alpha$ are removed from SED fitting analysis.}
\end{deluxetable}

\subsection{Lyman-Continuum Flux}
We used F150LP and F336W filters to detect LyC fluxes depending on galaxy redshifts. For LyC photometry, we defined apertures separately based on rest-frame far-UV (FUV) images that trace confined star-forming regions. We used the \texttt{Segmentation} subpackage in {\sc PHOTUTILS} to generate segmentation maps from the FUV images. We performed aperture photometry with the segmentation maps on the LyC-targeting HST filter images that are PSF-matched to their FUV counterparts.  Then, we measured LyC fluxes by summing fluxes within all segments within the tailored boundaries that we used in our broadband photometry in Section 2.4.  We investigated the LyC filter imaging (either F150LP or F336W) to find any positive fluxes within the segmentation maps built with FUV imaging, and we did not find significant ($>$2$\sigma$) signals from our targeted sources.  Additionally, we performed photometry on unconvolved LyC images to check if any LyC signals were smoothed out, which resulted in no ($>$2$\sigma$) detections as well.

Instead, we place tight upper limits of the LyC escape fraction by deriving realistic photometric uncertainties. The photometric errors in the LyC filter imaging were empirically derived by performing photometry using 100 random apertures of the same size and shape. We first performed aperture photometry (with the same segmentation maps) from 100 random pointings around the sources and took the standard deviation of a sigma-clipped sample, removing 5$\sigma$ outliers of the 100 random-aperture photometry. This provides empirically derived uncertainties of our customized photometry for the areas of interest where we expect to detect escaping LyC photons. We also corrected the Galactic extinction for the LyC photometry as described in Section 2.4. We take the estimated errors as the upper limits of the LyC fluxes for the rest of the analyses. The measured upper limits of LyC fluxes are listed in the 3rd column in Table \ref{tab:fesc}.

\begin{deluxetable*}{ccccccc}
\label{tab:properties}
\tabletypesize{\footnotesize}
\tablecaption{Physical Properties of Lensed Dwarf Galaxy Sample} 
\tablehead{
\colhead{ID} & \colhead{$\beta_{\rm UV}$} &  \colhead{$\beta_{\rm 1550}$}  & \colhead{$M_{\rm UV,uncor}$} & \colhead{$M_{\rm UV,cor}$$^{\dagger\dagger}$} & \colhead{Log(M$_{\rm star}/$M$_\odot$)$^{\dagger\dagger}$} & \colhead{$E(B-V)$}   \\
\colhead{(1)}  & \colhead{(2)} & \colhead{(3)} & \colhead{(4)} & \colhead{(5)} & \colhead{(6)} & \colhead{(7)}
}
\startdata
{SDSS J0851+3331-B} & {$-1.73\pm0.28$}  & {$-1.64\pm0.16$} & {$-20.45\pm0.29$}& {$-15.94^{+0.38}_{-0.92}$} & {$7.44^{+0.43}_{-0.08}$} & {$0.11^{+0.06}_{-0.05}$}  \\
{SDSS J0915+3826-A} & {$-2.10\pm0.07$}  & {$-2.06\pm0.08$} & {$-22.70\pm0.06$}& {$-17.92^{+0.35}_{-0.43}$} & {$9.14^{+0.15}_{-0.14}$} & {$0.02^{+0.02}_{-0.01}$} \\
{SDSS J1138+2754-A} & {$-1.97\pm0.22$}  & {$-1.93\pm0.26$} & {$-21.60\pm0.19$} & {$-19.65^{+0.19}_{-0.19}$} & {$10.11^{+0.00}_{-0.03}$} & {$0.06^{+0.07}_{-0.04}$} \\
{SDSS J1138+2754-C} & {$-2.06\pm0.40$}  & {$-2.00\pm0.46$} & {$-20.11\pm0.44$}& {$-14.37^{+1.16}_{-1.85}$} & {$7.82^{+0.56}_{-0.62}$} & {$0.05^{+0.10}_{-0.04}$} \\
{SDSS J1439+1208-A} & {$-2.23\pm0.06$}  & {$-2.22\pm0.07$} & {$-22.30\pm0.17$} & {$-15.65^{+0.34}_{-0.62}$} & {$7.96^{+0.18}_{-0.46}$} & {$0.01^{+0.02}_{-0.01}$} \\
{M0329-2556}        & {$-2.00\pm0.05$}  & {$-1.94\pm0.07$} & {$-20.20\pm0.03$}& {$-18.37^{+0.09}_{-0.11}$} & {$8.82^{+0.13}_{-0.11}$} & {$0.06^{+0.02}_{-0.02}$}  \\
{M1115-2511}        & {$-1.95\pm0.05$}  & {$-1.71\pm0.06$} & {$-20.48\pm0.02$}& {$-18.22^{+0.11}_{-0.08}$} & {$8.86^{+0.09}_{-0.11}$} & {$0.12^{+0.01}_{-0.02}$}  \\
{M1206-4340}        & {$-2.25\pm0.02$}  & {$-2.15\pm0.03$} & {$-22.88\pm0.01$}& {$-21.11^{+0.05}_{-0.08}$} & {$8.89^{+0.09}_{-0.13}$} & {$0.03^{+0.01}_{-0.01}$} 
\enddata
\tablecomments{Col.(1) Object ID, (2) Rest-UV slope measured from a best-fit SED at a rest-frame wavelength range of 1300\AA--2600\AA, (3) Rest-UV slope from a photometric power-law fit, (4) Absolute UV magnitude, (5) Magnification-corrected absolute UV magnitude, (6) Magnification-corrected stellar mass, (7) Dust Reddening}
\tablenotetext{}{
$^{\dagger\dagger}$\footnotesize Errors include magnification uncertainties.
}
\end{deluxetable*}

\begin{figure*}[th]
\centering
\includegraphics[width=1.0\textwidth]{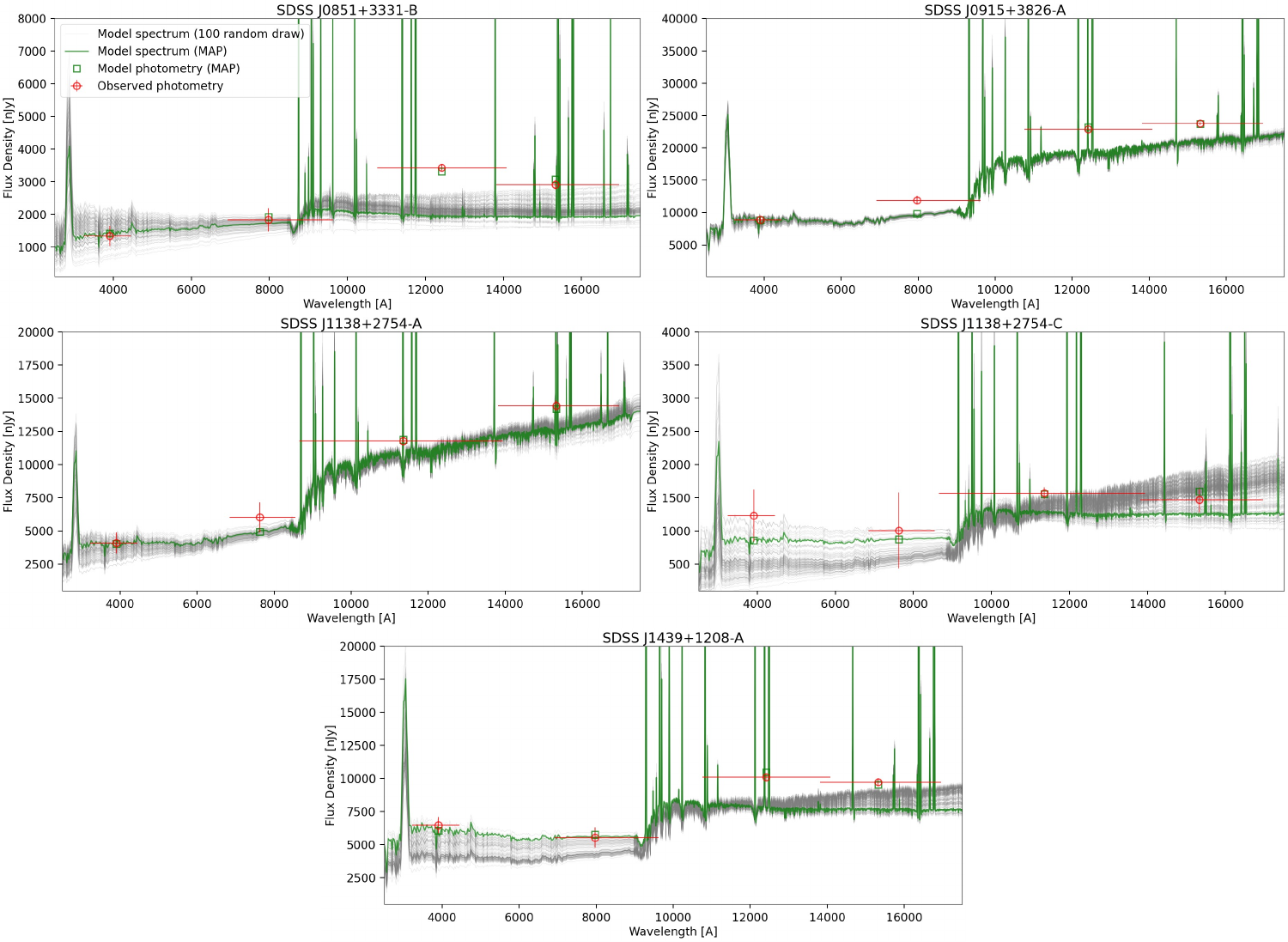}
\caption{Best-fit SEDs of SGAS targets from our {\sc Prospector} SED fitting analysis. The best-fit SEDs are green (MAP; maximum a posteriori), with 100 randomly drawn SEDs from MCMC chains (grey). The observed HST photometric data points are noted with open circles.} 
\label{fig:sed_sgas}
\end{figure*}

\begin{figure*}[th]
\centering
\includegraphics[width=1.0\textwidth]{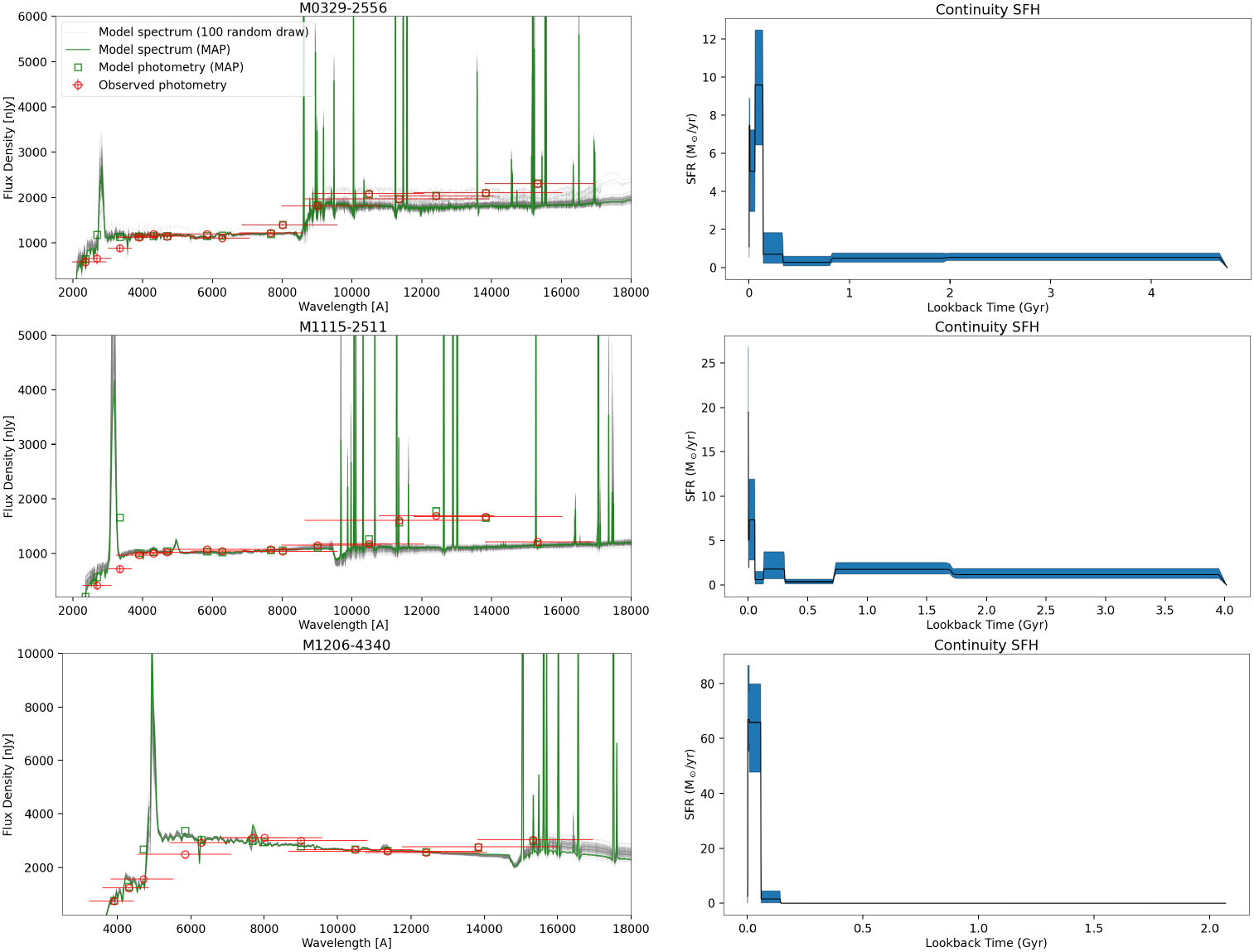}
\caption{(Left) Best-fit SEDs of CLASH targets from our {\sc Prospector} SED fitting analysis. The best-fit SEDs are green, with 100 randomly drawn SEDs from MCMC chains (grey). The observed HST photometric data points are noted with open circles. (Right) Star-formation histories recovered from SED modeling with nonparametric SFHs in {\sc Prospector}. The SFHs are modeled in eight temporal bins in lookback time.} 
\label{fig:sed_clash}
\end{figure*}

\section{Analysis}
\subsection{Galaxy Properties and Star-formation Histories}
To derive galaxy physical properties and study their star formation histories (SFHs), we perform galaxy spectral energy distribution (SED) fitting with {\sc Prospector} \citep{Leja2017a, Johnson2021a} for our sample. {\sc Prospector} uses the Flexible Stellar Population Synthesis (FSPS) models \citep{Conroy2009a} with the Modules for Experiments in Stellar Astrophysics Isochornes and Stellar Tracks \citep[MIST;][]{Choi2016a} and {\sc CLOUDY} \citep{Ferland2013a} to include the nebular emission lines as described in \cite{Byler2017a}. {\sc Prospector} is capable of modeling SEDs with flexible (so-called "nonparametric") SFHs. Specifically, {\sc Prospector} models nonparametric SFHs with binned SFHs, described as collective piece-wise constant star-formation rates (SFRs) of predefined temporal bins. Although this adds, in general, more free parameters to fit (i.e., SFRs in individual temporal bins), we have 16-band photometric data points in the CLASH fields, which enables SED fitting analysis with nonparametric SFHs. However, only four HST photometric bands are available for SGAS targets. Thus, we adopt a parametric SFH in SED fitting to reduce free parameters to fit for SGAS galaxies. Our targets are spectroscopically confirmed; thus, we fix the redshifts to the spectroscopic redshifts. We model dust attenuation by adopting the \cite{Calzetti2000a} dust attenuation curve. In {\sc Prospector}, the IGM attenuation is modeled following \cite{Madau1995a} with a scaling parameter of the IGM attenuation curve. The IGM scaling parameter accounts for line-of-sight variations in the total IGM opacity.

The free parameters in SED fitting for CLASH targets include dust attenuation, stellar and gas metallicities, an ionization parameter, and IGM attenuation in addition to eight temporal bins to model nonparametric SFHs. We adopt top-hat priors to the free parameters, allowing the $V$-band optical depth ($\tau_V$) from 0 to 4 in dust attenuation, metallicities in the range of $-2.0\leq {\rm log}(Z/Z_\odot) \leq2.0$ (gas metallicity is tied to stellar metallicity), the ionization parameter in the range of $-4.0\leq {\rm log}(U) \leq-1.0$. The SFHs are modeled in eight temporal bins that include the two most recent temporal bins of [0, 3] Myr and [3, 10] Myr in addition to six evenly spaced bins in logarithmic time between 10 Myr to the age of the universe at a source redshift.  For IGM attenuation, we allow the IGM scaling parameter, that accounts for line-of-sight variations of the total IGM opacity, from 0 to 2, with a Gaussian prior centered at 1.

For SGAS targets where we have far fewer photometric bands, we fix the metallicity to ${\rm log}(Z/Z_\odot)=-0.7$, the ionization parameter to ${\rm log}(U)=-2$, the IGM scaling parameter to the unity. The free parameters are dust attenuation ($0\leq\tau_V\leq2$), stellar population age (limited by the age of the universe at a source redshift), and a $\tau$ value in the delayed-$\tau$ SFH with the range of $-1.0<{\rm log}(\tau)<10.0$. We allow stellar mass in the range of $5\leq {\rm log}(M_*/M_\odot) \leq12$. 

The rest-frame UV slope ($\beta_{\rm UV}$) is defined as a galaxy UV spectral index ($F_{\lambda}\propto \lambda^{\beta}$ or $F_{\nu}\propto \lambda^{\beta+2}$). We derived the UV slope from best-fit SED models over 1300\,--\,2600\AA\ which use all available photometric data points. To compare to the local LyC sample (see Section 4), we separately derived UV slopes centered at 1550\AA\ over a spectral window of 1300\,--\,1800\AA, as defined in \cite{Chisholm2022a}.  UV slopes can be measured directly as well from photometric power-law fitting. However, we have limited rest-UV coverage for SGAS targets; thus, we rely on SED-based measurements. It has also been suggested that SED-derived UV slopes are less likely biased than those from photometric power-law fitting except for the cases of extremely blue rest-UV slopes ($\beta<-3$) unavailable in SED templates \citep[e.g.,][]{Finkelstein2012a, Larson2023b, Morales2023a}. 

The best-fit SEDs of our SGAS targets are displayed in Figure \ref{fig:sed_sgas}. We also present the best-fit SEDs of the CLASH targets and their reconstructed nonparametric SFHs in Figure \ref{fig:sed_clash}.  The key quantities derived from SED fitting and the rest-UV slopes are listed in Table \ref{tab:properties}.

\subsection{Lensing Models}
Our targets are highly magnified with gravitational lensing caused by foreground clusters of galaxies.  We obtained the latest available lensing models of the seven foreground clusters behind which our sources are located. The lensing models of SGAS clusters are available in \cite{Sharon2020a}. We used the lens models in \cite{Zitrin2015a} for CLASH clusters. Using the cluster lensing models, we estimated the magnification factors at the source centers as listed in Table \ref{tab:targets}.  We corrected the UV absolute magnitudes and stellar masses by dividing them by the estimated magnifications to discuss the intrinsic properties.

\begin{deluxetable*}{cccccccc}
\label{tab:fesc}
\tabletypesize{\footnotesize}
\tablecaption{LyC Escape Properties of Lensed Dwarf Galaxy Sample$^{**}$} 
\tablehead{
\colhead{ID} & \colhead{$\lambda_{\rm LyC}$ (\AA)} & \colhead{$F_{\rm LyC}$ ($\mu$Jy)} & \colhead{$F_{\rm LyC}/F_{\rm 1500}$$^{\dagger}$} & \colhead{$T_{\rm IGM}$} & $(L_{1500}/L_{\rm LyC})_{\rm stel}$ & \colhead{$f_{\rm esc,rel}$}& \colhead{$f_{\rm esc,abs}$}  \\
\colhead{(1)}  & \colhead{(2)} & \colhead{(3)} & \colhead{(4)} & \colhead{(5)} & \colhead{(6)} & \colhead{(7)} & \colhead{(8)}
}
\startdata
{SDSS J0851+3331-B} & {$685$} & {$<0.011$} & {$<0.013$} & {$0.61$} & {$8$} & {$<0.106$} & {$<0.046$}  \\
{SDSS J0915+3826-A} & {$642$} & {$<0.044$} & {$<0.007$} & {$0.54$} & {$8$} & {$<0.056$} & {$<0.048$}  \\
{SDSS J1138+2754-A} & {$688$} & {$<0.023$}  & {$<0.005$} & {$0.62$} & {$8$} & {$<0.043$} & {$<0.027$}  \\
{SDSS J1138+2754-C} & {$654$} & {$<0.009$}  & {$<0.029$} & {$0.56$} & {$8$} & {$<0.222$} & {$<0.152$}  \\
{SDSS J1439+1208-A} & {$644$} & {$<0.008$}  & {$<0.008$} & {$0.54$} & {$8$} & {$<0.062$} & {$<0.057$}  \\
{M0329-2556}        & {$694$} & {$<0.010$}  & {$<0.014$} & {$0.63$} & {$8$} & {$<0.112$} & {$<0.071$}  \\
{M1115-2511}        & {$618$} & {$<0.005$}  & {$<0.009$} & {$0.50$} & {$8$} & {$<0.075$} & {$<0.030$}  \\
{M1206-4340}        & {$831$} & {$<0.028$}  & {$<0.030$} & {$0.32$} & {$6$} & {$<0.178$} & {$<0.142$}
\enddata
\tablecomments{Col.(1) Object ID, (2) Rest-frame central wavelength of LyC HST-filter-band, (3) Upper limits of LyC fluxes, (4) Flux ratio between the LyC-band flux and rest-UV flux at 1500\AA, (5) Mean IGM transmission at a source redshift at $\lambda_{\rm LyC}$ based on \cite{Inoue2014a}, (6) Intrinsic stellar Lyman break adopted in Eq. \ref{eqn:fesc}, (7) Relative LyC escape fraction following Eq. \ref{eqn:fesc}, (8) Absolute LyC escape fraction, UV-dust-attenuation ($A_{\rm UV}$) corrected as $f_{\rm esc,rel} \times 10^{-0.4 A_{\rm UV}}$ }
\tablenotetext{}{
$^{**}$\footnotesize 1$\sigma$ upper limits are provided.\\
$^{\dagger}$\footnotesize Corrected for IGM attenuation.
}
\end{deluxetable*}

\subsection{LyC Escape Fraction}
We measure the upper limits of LyC fluxes from LyC detection bands as described in Section 2.5. The escape fraction of LyC photons is estimated as the ratio of the measured LyC fluxes to the intrinsic LyC fluxes. As we do not find significant LyC fluxes from our observations, we use the upper limits of LyC fluxes to calculate the upper limits of the LyC escape fractions. 

The relative escape fractions of LyC fluxes are defined following previous studies \citep[e.g.,][]{Steidel2001a, Siana2007a, Siana2010a}:
\begin{equation}
f_{\rm esc,rel} = \frac{(L_{1500}/L_{\rm LyC})_{\rm stel}}{(F_{1500}/F_{\rm LyC})_{\rm obs}}\,{\rm exp(\tau_{IGM})},   
\label{eqn:fesc}
\end{equation}
where $(F_{1500}/F_{\rm LyC})_{\rm obs}$ is the observed flux ratio between the flux density ($F_{\nu}$) at 1500\AA\ ($F_{1500}$) and in the LyC wavelength ($F_{\rm LyC}$; $<$900\AA). The 1500\AA\ fluxes ($F_{1500}$) are taken average in a range of 1450 -- 1550\AA\ in the rest frame in the best-fit SEDs obtained from the {\sc Prospector} runs. $(L_{1500}/L_{\rm LyC})_{\rm stel}$ represents the intrinsic flux ratio from stellar populations modeling.  Consistent with most of the literature, we adopt a $(L_{1500}/L_{\rm LyC})_{\rm stel}$ value of 8 at $\sim$700\AA\ and 6 at $\sim$900\AA\ as assumed in \cite{Siana2007a}. The intrinsic flux ratio could be much larger depending on star-formation histories, particularly if galaxies have a single epoch of bursty star formation in the past as the contribution from massive stars decreases rapidly \citep{Siana2007a, Dominguez2015a, Alavi2020a}.  However, our targets were selected as having blue UV slopes indicating recent star formation with young stellar ages; adopting the standard value of 6 -- 8 for actively star-forming galaxies thus seems justified. The correction term for IGM attenuation by neutral hydrogen is added as ${\rm exp(\tau_{IGM})}$ where $\tau_{\rm IGM}$ is the IGM opacity at the LyC wavelength ($\sim$620 -- 830\AA). We use the IGM attenuation model \citep{Inoue2014a} and obtain the redshift-dependent mean IGM transmissions of ${\rm exp(-\tau_{IGM})}\sim0.6$ at $z\sim1.3$ and $\sim0.3$ at $z\sim3$.  This IGM transmission value at $z\sim3$ agrees with the empirically measured IGM transmission at $z\sim3$ in \cite{Steidel2018a}. However, we caution that the sight line variation of IGM transmission could be significant at such high redshifts, which may affect a resulting upper limit of LyC escape up to around a factor of two. We present the estimated $f_{\rm esc,rel}$ upper limits. 

The definition of $f_{\rm esc,rel}$ can be related to the absolute escape fraction ($f_{\rm esc,abs}$) that accounts for dust attenuation of UV photons ($A_{\rm UV}$): $f_{\rm esc,abs} = f_{\rm esc,rel} \times 10^{-0.4 A_{\rm UV}}$.  Thus, the absolute escape fractions can be lower than the relative escape fractions as dust attenuation becomes significant \citep[e.g.,][]{Shapley2016a, Kerutt2023a}. We calculated the upper limits of the absolute LyC escape fraction using the \cite{Calzetti2000a} attenuation curve. 

We summarize all LyC-escape-related properties of our sources in Table \ref{tab:fesc}. 

\section{Discussion}

\begin{figure*}[ht]
\centering
\includegraphics[width=\textwidth]{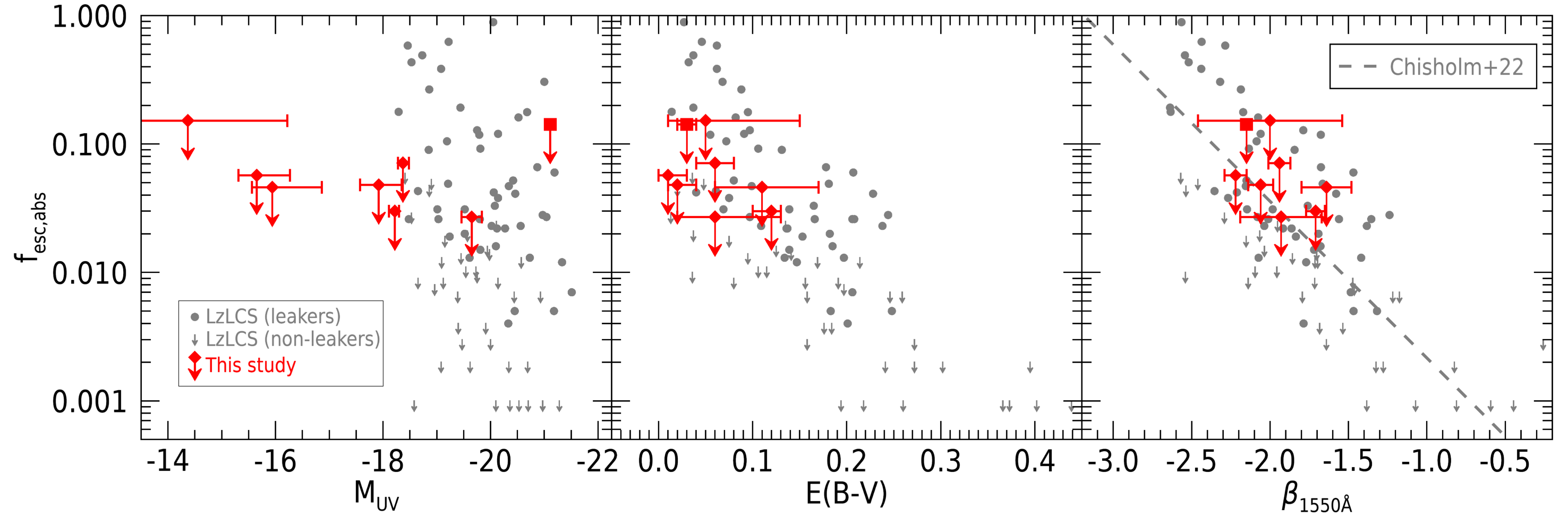}
\caption{Relative escape fraction of LyC photons \textit{vs.} absolute UV magnitude (left), dust reddening (middle), and UV $\beta$ slope (right). We present the 1$\sigma$ upper limits of LyC escape fractions as red arrows. The square symbol shows the $z\approx3$ target, and the other targets are marked as diamonds. We include the known $z\sim0.3$ sample of LyC leakers in LzLCS+ \citep{Chisholm2022a, Saldana-Lopez2022a}. The significant LyC leakers are shown as filled circles, and the grey arrows represent non-leakers in the LzLCS+ sample.} 
\label{fig:fesc}
\end{figure*}

\subsection{LyC Escape with Galaxy Properties}
To place the upper limits on LyC escape in context, it is useful to compare the properties of galaxies in our sample to other samples of galaxies that have been used to study LyC escape. In Figure \ref{fig:fesc}, we present the upper limits of the LyC escape fraction ($f_{\rm esc,abs}$) of our targets in terms of UV luminosity, dust attenuation, and rest-UV slope. We compare our measurements to the known $z\sim0.3$ sample in LzLCS+ \citep{Chisholm2022a, Saldana-Lopez2022a}, which combines the 23 LyC emitters \citep{Izotov2016a, Izotov2016b, Izotov2018a, Izotov2018b, Wang2019a, Izotov2021a} and the LzLCS sample \citep{Flury2022a, Flury2022b}.

Compared to the literature studies, we explore a much fainter regime down to $M_{\rm UV}\sim-14$ thanks to gravitational lensing magnification, as shown in the left panel of Figure \ref{fig:fesc}. Interestingly, we constrain $f_{\rm esc}$ upper limits from three faint sources ($M_{\rm UV}>-16$), that are comparable to those from relatively brighter targets ($M_{\rm UV}\lesssim-18$). This mainly suggests no evidence of the increased escape fractions of LyC photons from intrinsically UV faint sources although this must be further explored with additional observations. 

Also, our SED fitting analysis suggests low dust content in these galaxies with $E(B-V)\lesssim0.1$ (middle panel). There is a significant increasing trend of $f_{\rm esc}$ with lower dust content seen from the LzLCS+ sample. Although we do not find evidence of LyC photon escape from our sources, we place $f_{\rm esc}$ upper limits around the lower bound of the LyC-leaker distribution of LzLCS+, yet compatible with findings from the low-redshift sample. 

In the last panel, we present the $f_{\rm esc}$ upper limits versus the rest-UV slope at 1550\AA\ ($\beta_{\rm 1550}$). \cite{Chisholm2022a} find a correlation between $f_{\rm esc}$ and $\beta_{\rm 1550}$ from the LzLCS+ sample where they find significant LyC leakers with blue UV slopes ($\beta<-1.5$).  Our targets exhibit blue UV slopes similar to the LyC leakers in the LzLCS+ sample.  Despite non-detections of LyC fluxes, similar to what we find for $E(B-V)$, our results on the $f_{\rm esc}$ upper limits lie along with the correlation suggested in \cite{Chisholm2022a}.

\begin{figure*}[ht]
\centering
\includegraphics[width=\textwidth]{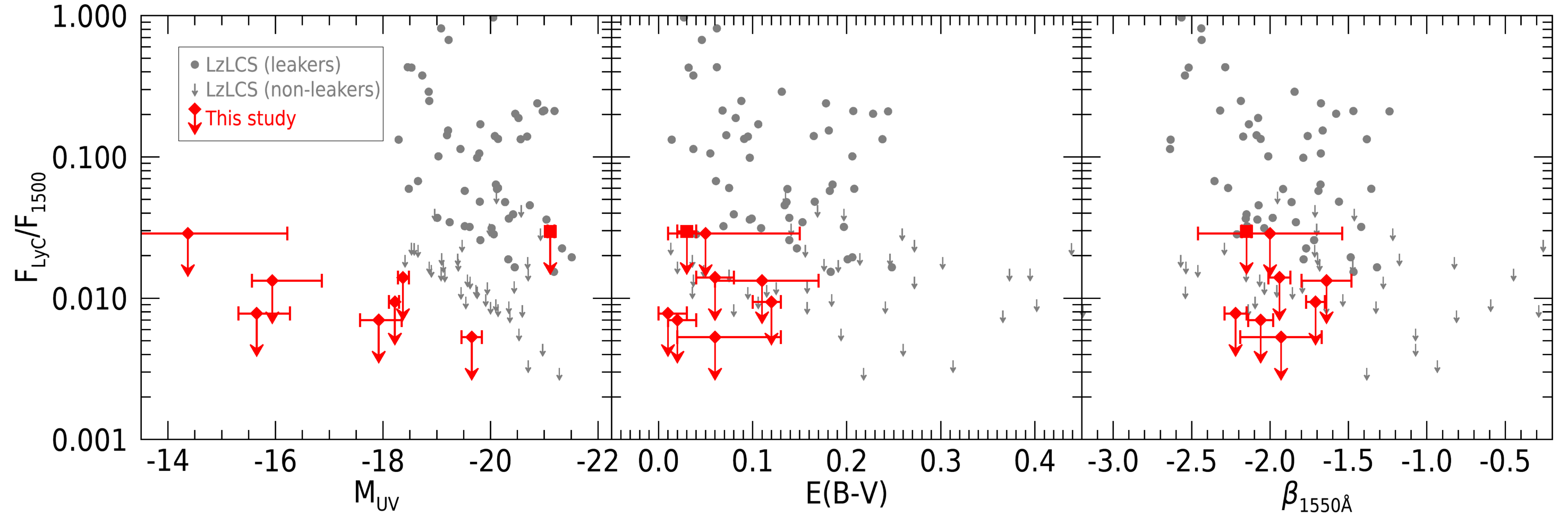}
\caption{Similar to Figure \ref{fig:fesc}, but showing the ratio of LyC fluxes to UV fluxes at 1500\AA\ as a function of absolute UV magnitude (left), dust reddening (middle), and UV $\beta$ slope (right).} 
\label{fig:fescobs}
\end{figure*}

Separately, for comparison on more direct observables, we derived the upper limits for the ratio of observed LyC flux to FUV continuum fluxes ($F_{\rm LyC}/F_{1500}$), that are corrected for IGM attenuation, dropping the assumptions of the intrinsic flux ratios ($L_{1500}/L_{\rm LyC}$). The $F_{\rm LyC}/F_{1500}$ values are listed in the 4th column in Table \ref{tab:fesc}. For a comparison, UV continuum fluxes ($F_{1500}$) of the LzLCS+ sample are converted from their $M_{\rm 1500}$ values \citep{Flury2022b}. In Figure \ref{fig:fescobs}, we show the upper limits of $F_{\rm LyC}/F_{1500}$ of our targets as well as the LzLCS+ sample with the same physical properties to Figure \ref{fig:fesc}.  In this analysis, the galaxies in our sample tend to depart from the trend for LzLCS+ LyC leakers with low \textit{E(B-V)} or very blue UV slopes to have larger $F_{\rm LyC}/F_{1500}$.  This result could be further tested even for the current sample by obtaining more complete multi-wavelength photometry for the SGAS sample galaxies to understand better their SFHs and emission-line properties to measure emission-line-based $f_{\rm esc}$.\citep[e.g.,][]{Rutkowski2016a, Fletcher2019a, Alavi2020a}.

\subsection{Interpretations of the Non-detections}
Our targets pose as potential LyC leakers with galaxy properties similar to those of confirmed LyC leakers from the literature. The low dust contents and blue UV slopes found in our targets suggest that the galaxies are young and in an active star-forming phase. As described in Section 3.1, we further examine SFHs of our CLASH targets thanks to the capability of building nonparametric SFHs in {\sc Prospector}. The reconstructed SFHs of CLASH targets are displayed in the right panels in Figure \ref{fig:sed_clash}. All three targets show recent bursty star formation within $\sim100$ Myrs in the past. This implies young stellar ages and is consistent with blue UV slopes.  Such galaxy properties suggest that the galaxies are likely leaking LyC photons. Particularly, higher LyC escape fractions in faint and less massive galaxies have been suggested in many theoretical studies \citep[e.g.,][]{Wise2014a, Paardekooper2015a, Xu2016a, Anderson2017a}. However, our findings do not support the enhanced escape fraction of LyC photons from faint sources. 

This seems puzzling at a glance, but it is understandable if we consider scenarios of highly anisotropic geometry of LyC escape with varying ISM conditions depending on sight lines. This has been suggested in theoretical studies \citep[e.g.,][]{Cen2015a, Ma2015a, Paardekooper2015a, Witten2023a} where they present significant line-of-sight variations in the escape fraction of LyC photons. Observationally, \cite{Kim2023a} reported significant spatial variations of LyC leakage from their spatially-resolved analysis of HST imaging of the Sunburst Arc: a strongly lensed LyC emitter at $z = 2.37$. Such an anisotropic nature of LyC escape is consistent, in general, with the relatively large scatter found in relations between individual galaxy physical properties and the LyC escape fraction \citep[e.g.,][]{Izotov2018a, Chisholm2022a, Flury2022b}. Indeed, similar to what is reported for the Sunburst Arc, a possible anisotropy of LyC escape has been suggested by the complex ionized gas kinematics (i.e. asymmetry of line profiles) of LzLCS LyC leakers \citep{Amorin2024a}.

Alternatively, the non-detections of escaping LyC fluxes in our study may simply indicate no significant increase in the LyC escape fraction for fainter galaxies.  A turnover of increasing LyC escape fraction with lower stellar mass has been discussed due to less efficient star formation in low-mass-end populations with $M_*\lesssim10^{7}M_{\odot}$ \citep{Ma2020a, Rosdahl2022a} although our targets are suggested to be slightly more massive than this threshold.  Interestingly, such low luminosity galaxies have been observed in Ly$\alpha$, the emission of which is dependent primarily on \ion{H}{i} covering/column density and dust, in a qualitatively similar way to LyC \citep[e.g.,][for a review]{Verhamme2015a, Hayes2015a}.  Indeed, Ly$\alpha$ escape and LyC escape correlate well in some existing samples \citep{Izotov2020a, Flury2022b, Pahl2021a, Steidel2018a}.  Some such low luminosity galaxies were targeted early with HST: specifically IZw18 ($M_{\rm UV}\sim-15.2$), SBS\,0335--052E ($M_{\rm UV}\sim-16.2$), Tol\,65 ($M_{\rm UV}\sim-14.6$), and SBS\,1415+437 ($M_{\rm UV}\sim-15.4$) overlap with the low luminosity end of our lensed sample. These galaxies all show very strong Ly$\alpha$ absorption, with EWs $\lesssim -25$~\AA\ \citep{Kunth1994a, Thuan1997a, James2014a}, but show very strong H$\alpha$ emission implying vigorous levels of ongoing star formation. The damped Ly$\alpha$ with no hints of emission implies the starbursts are covered by thick layers of neutral hydrogen with column densities $\sim 2\times 10^{21}$~cm$^{-2}$, with similar phenomena observed in numerous galaxies from the CLASSY sample \citep{Hu2023a}. These galaxies mostly show high atomic gas mass fractions, and we have previously speculated that the very youthful star formation in these galaxies has not yet accelerated or disrupted the absorbing material. At least along the line of sight, these galaxies must almost inevitably show zero LyC emission if the observation could be made. Broadly speaking, we speculate that similar phenomena could be responsible for the lack of LyC emission we report in our very faint lensed dwarfs. For example, the MUSE spectra for some of our targets (M0329-2556 and M1206-4340) present either strong damped Ly$\alpha$ feature or abundant Balmer and metal absorption lines while ground-based spectra of our SGAS targets are relatively shallow, limiting further spectral information but their redshift confirmations \citep{Caminha2017a, Richard2021a}.

Lastly, it is worth noting that $f_{\rm esc}$ measurements from imaging may be more complicated to interpret than what is measured from spectra. It relies on, for example, the exact shape of a filter transmission curve \citep[e.g., see discussion in][]{Steidel2018a}. A further comparison of LyC measurements from photometry to those from spectra will help in understanding potential biases.

\section{Summary}
We present HST imaging observations with the ACS/SBC F150LP and WFC3/UVIS F336W filters, aiming to detect LyC photons from eight highly magnified sources at $z=1.3$\,--\,$3.0$ with blue UV slopes ($\beta<-1.7$). Our targets include intrinsically faint galaxies in their rest-frame UV, down to $M_{\rm UV}$\,$\sim$\,$-14$.  We perform photometry with custom-defined apertures to derive reliable measurements of LyC escape dealing with complex morphologies in our lensed targets. Our findings are summarized as follows. 

\begin{enumerate}
    \item{With the PSF-matched HST imaging data, we perform custom-defined aperture photometry to ensure the inclusion of extended source boundaries of our lensed sources.}
    \item{We perform SED fitting with {\sc Prospector} to derive key physical quantities, including absolute UV luminosity, dust reddening, and UV $\beta$ slope. Our targets pose as potential LyC leakers with low dust content, faint in UV, and blue in UV color. We model SEDs with nonparametric SFHs for three CLASH targets, which recovers bursty star formation within the recent $\lesssim$100 Myrs in these galaxies.}
    \item{We derive tight upper limits on the escape fraction of LyC photons by estimating empirically derived photometry errors. From our observations, we report no significant ($>$\,$2\sigma$) detections of LyC fluxes, placing 1$\sigma$ upper limits at 3 -- 15\% of the LyC escape fractions ($f_{\rm esc,abs}$).}
    \item{We constrain $f_{\rm esc}$ upper limits from three faint sources ($M_{\rm UV}>-16$), that are comparable to those from relatively brighter targets ($M_{\rm UV}\lesssim-18$). Albeit with a relatively small sample -- and with relatively large uncertainties in some of the magnification estimates -- our results do not show the expected increase of LyC escape fraction from very-low luminosity galaxies.} 
    \item{Despite the non-detections of LyC fluxes, the $f_{\rm esc,abs}$ upper limits of our sources are in general compatible with those of the LyC-leakers in the LzLCS+ sample that share similar physical properties with our targets (i.e. less dusty and blue in rest UV color). However, the galaxies in our sample tend to depart from the trend for the low-redshift LyC leakers with low $E(B-V)$ or very blue UV slopes ($\beta$) to have larger $F_{\rm LyC}/F_{1500}$. This result could be further tested even for the current sample by obtaining more complete multi-wavelength photometry for the SGAS sample galaxies to understand better their SFHs and emission-line properties to study emission-line-based $f_{\rm esc}$.}
\end{enumerate}

UV faint galaxies have been suspected to be the main drivers of reionization. Understanding their contribution to ionizing photon supplies to the IGM can be critical to constrain the detailed reionization processes. Recent JWST observations suggest that UV faint galaxies are likely efficient in producing ionizing photons \citep[e.g.,][]{Atek2023a, Simmonds2024a}. However, unknown LyC escape makes it challenging to diagnose the actual capability of these faint galaxies in supplying ionizing photons to the IGM. In particular, we find no significant enhancement of LyC escape in UV faint galaxies. Increasing the sample in these faint and low-mass galaxies is critical in future LyC observations to understand their relative contributions to reionization.

\begin{acknowledgments}
This work is primarily based on observations taken by the HST Cycle 25 program (GO 15271) with the NASA/ESA HST, which is operated by the Association of Universities for Research in Astronomy, Inc., under NASA contract NAS5-26555. Support for this work was provided by NASA through a grant program ID HST-GO-15271 from the Space Telescope Science Institute, which is operated by the Association of Universities for Research in Astronomy, Inc., under NASA contract NAS5-26555. Some of the data presented in this paper were obtained from the Multimission Archive at the Space Telescope Science Institute (MAST). These data are associated with programs GO-13003 (SGAS):\dataset[10.17909/t9-cqtj-y020]{\doi{10.17909/t9-cqtj-y020}}, GO-12065 (CLASH):\dataset[10.17909/T90W2B]{\doi{10.17909/T90W2B}}, GO-11387 (NGC6681), HA is supported by the Centre National d'Etudes Spatiales (CNES).
\end{acknowledgments}

\vspace{5mm}
\facilities{\textit{HST} (ACS and WFC3)}

\software{astropy \citep{Astropy2018a},
          IDL Astronomy Library: \url{idlastro.gsfc.nasa.gov} \citep{Landsman1993a},
          matplotlib \citep{Hunter2007a},
          NumPy \citep{Harris2020a}, 
          SciPy \citep{Virtanens2020a},
          Photutils \citep{Bradley2023a},
          DrizzlePac \citep{Fruchter2010a}
}

\appendix
\section{Building the Point-Spread Function Model of F150LP}
\begin{figure}[th]
\centering
\includegraphics[width=0.5\columnwidth]{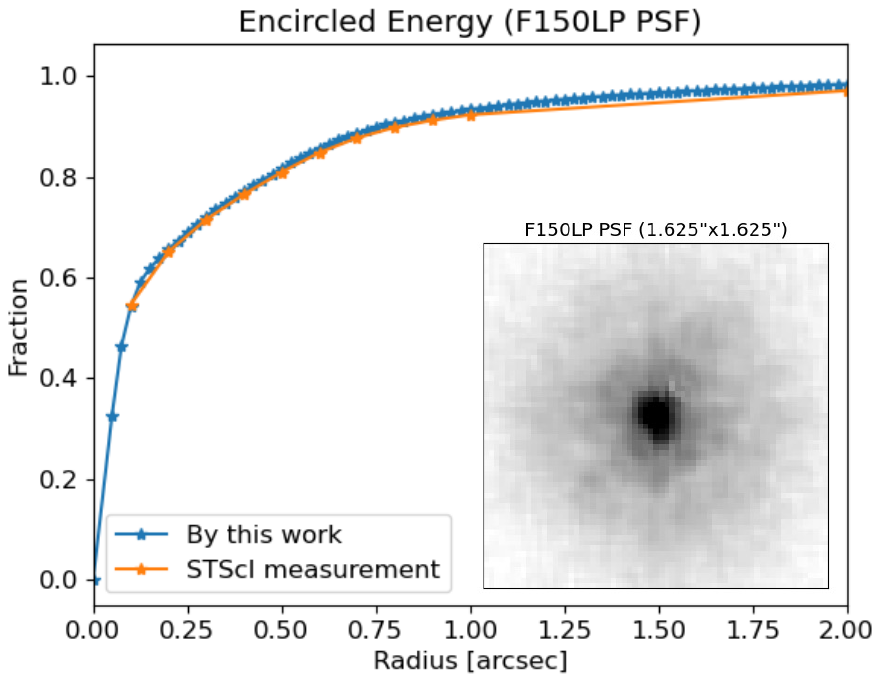}
\caption{The encircled energy curve of the F150LP PSF model (blue), compared to the STScI measurement (orange). The inset panel shows the constructed PSF model of F150LP within a 1\farcs625$\times$1\farcs625 box.} 
\label{fig:psf}
\end{figure}

We built the PSF model of F150LP as described in Section 2.4. The measured encircled energy curve from the PSF model agrees well with the measurement in the STScI webpage, as shown in Figure \ref{fig:psf}. The constructed F150LP PSF model is presented in the inset panel of Figure \ref{fig:psf}.

\bibliographystyle{aasjournal}
\bibliography{references}

\begin{thebibliography}{}
\expandafter\ifx\csname natexlab\endcsname\relax\def\natexlab#1{#1}\fi
\providecommand{\url}[1]{\href{#1}{#1}}
\providecommand{\dodoi}[1]{doi:~\href{http://doi.org/#1}{\nolinkurl{#1}}}
\providecommand{\doeprint}[1]{\href{http://ascl.net/#1}{\nolinkurl{http://ascl.net/#1}}}
\providecommand{\doarXiv}[1]{\href{https://arxiv.org/abs/#1}{\nolinkurl{https://arxiv.org/abs/#1}}}

\bibitem[{{Alavi} {et~al.}(2020){Alavi}, {Colbert}, {Teplitz}, {Siana},
  {Scarlata}, {Rutkowski}, {Mehta}, {Henry}, {Dai}, {Haardt}, \&
  {Bagley}}]{Alavi2020a}
{Alavi}, A., {Colbert}, J., {Teplitz}, H.~I., {et~al.} 2020, \apj, 904, 59,
  \dodoi{10.3847/1538-4357/abbd43}

\bibitem[{{Amor{\'\i}n} {et~al.}(2014){Amor{\'\i}n}, {Grazian}, {Castellano},
  {Pentericci}, {Fontana}, {Sommariva}, {van der Wel}, {Maseda}, \&
  {Merlin}}]{Amorin2014a}
{Amor{\'\i}n}, R., {Grazian}, A., {Castellano}, M., {et~al.} 2014, \apjl, 788,
  L4, \dodoi{10.1088/2041-8205/788/1/L4}

\bibitem[{{Amor{\'\i}n} {et~al.}(2024){Amor{\'\i}n},
  {Rodr{\'\i}guez-Henr{\'\i}quez}, {Fern{\'a}ndez}, {V{\'\i}lchez},
  {Marques-Chaves}, {Schaerer}, {Izotov}, {Firpo}, {Guseva}, {Jaskot},
  {Komarova}, {Mu{\~n}oz-Vergara}, {Oey}, {Bait}, {Carr}, {Chisholm},
  {Ferguson}, {Flury}, {Giavalisco}, {Hayes}, {Henry}, {Ji}, {King},
  {Leclercq}, {{\"O}stlin}, {Pentericci}, {Saldana-Lopez}, {Thuan},
  {Trebitsch}, {Wang}, {Worseck}, \& {Xu}}]{Amorin2024a}
{Amor{\'\i}n}, R.~O., {Rodr{\'\i}guez-Henr{\'\i}quez}, M., {Fern{\'a}ndez}, V.,
  {et~al.} 2024, arXiv e-prints, arXiv:2401.04278,
  \dodoi{10.48550/arXiv.2401.04278}

\bibitem[{{Anderson} {et~al.}(2017){Anderson}, {Governato}, {Karcher}, {Quinn},
  \& {Wadsley}}]{Anderson2017a}
{Anderson}, L., {Governato}, F., {Karcher}, M., {Quinn}, T., \& {Wadsley}, J.
  2017, \mnras, 468, 4077, \dodoi{10.1093/mnras/stx709}

\bibitem[{{Arrabal Haro} {et~al.}(2023{\natexlab{a}}){Arrabal Haro},
  {Dickinson}, {Finkelstein}, {Kartaltepe}, {Donnan}, {Burgarella}, {Carnall},
  {Cullen}, {Dunlop}, {Fern{\'a}ndez}, {Fujimoto}, {Jung}, {Krips}, {Larson},
  {Papovich}, {P{\'e}rez-Gonz{\'a}lez}, {Amor{\'\i}n}, {Bagley}, {Buat},
  {Casey}, {Chworowsky}, {Cohen}, {Ferguson}, {Giavalisco}, {Huertas-Company},
  {Hutchison}, {Kocevski}, {Koekemoer}, {Lucas}, {McLeod}, {McLure}, {Pirzkal},
  {Trump}, {Weiner}, {Wilkins}, \& {Zavala}}]{Arrabal-Haro2023a}
{Arrabal Haro}, P., {Dickinson}, M., {Finkelstein}, S.~L., {et~al.}
  2023{\natexlab{a}}, arXiv e-prints, arXiv:2303.15431,
  \dodoi{10.48550/arXiv.2303.15431}

\bibitem[{{Arrabal Haro} {et~al.}(2023{\natexlab{b}}){Arrabal Haro},
  {Dickinson}, {Finkelstein}, {Fujimoto}, {Fern{\'a}ndez}, {Kartaltepe},
  {Jung}, {Cole}, {Burgarella}, {Chworowsky}, {Hutchison}, {Morales},
  {Papovich}, {Simons}, {Amor{\'\i}n}, {Backhaus}, {Bagley}, {Bisigello},
  {Calabr{\`o}}, {Castellano}, {Cleri}, {Dav{\'e}}, {Dekel}, {Ferguson},
  {Fontana}, {Gawiser}, {Giavalisco}, {Harish}, {Hathi}, {Hirschmann},
  {Holwerda}, {Huertas-Company}, {Koekemoer}, {Larson}, {Lucas}, {Mobasher},
  {P{\'e}rez-Gonz{\'a}lez}, {Pirzkal}, {Rose}, {Santini}, {Trump}, {de la
  Vega}, {Wang}, {Weiner}, {Wilkins}, {Yang}, {Yung}, \&
  {Zavala}}]{Arrabal-Haro2023b}
---. 2023{\natexlab{b}}, arXiv e-prints, arXiv:2304.05378,
  \dodoi{10.48550/arXiv.2304.05378}

\bibitem[{{Atek} {et~al.}(2023){Atek}, {Labb{\'e}}, {Furtak}, {Chemerynska},
  {Fujimoto}, {Setton}, {Miller}, {Oesch}, {Bezanson}, {Price}, {Dayal},
  {Zitrin}, {Kokorev}, {Weaver}, {Brammer}, {van Dokkum}, {Williams}, {Cutler},
  {Feldmann}, {Fudamoto}, {Greene}, {Leja}, {Maseda}, {Muzzin}, {Pan},
  {Papovich}, {Nelson}, {Nanayakkara}, {Stark}, {Stefanon}, {Suess}, {Wang}, \&
  {Whitaker}}]{Atek2023a}
{Atek}, H., {Labb{\'e}}, I., {Furtak}, L.~J., {et~al.} 2023, arXiv e-prints,
  arXiv:2308.08540, \dodoi{10.48550/arXiv.2308.08540}

\bibitem[{{Bayliss} {et~al.}(2011){Bayliss}, {Gladders}, {Oguri}, {Hennawi},
  {Sharon}, {Koester}, \& {Dahle}}]{Bayliss2011a}
{Bayliss}, M.~B., {Gladders}, M.~D., {Oguri}, M., {et~al.} 2011, \apjl, 727,
  L26, \dodoi{10.1088/2041-8205/727/1/L26}

\bibitem[{{Begley} {et~al.}(2022){Begley}, {Cullen}, {McLure}, {Dunlop},
  {Hall}, {Carnall}, {Hamadouche}, {McLeod}, {Amor{\'\i}n}, {Calabr{\`o}},
  {Fontana}, {Fynbo}, {Guaita}, {Hathi}, {Hibon}, {Ji}, {Llerena},
  {Pentericci}, {Saldana-Lopez}, {Schaerer}, {Talia}, {Vanzella}, \&
  {Zamorani}}]{Begley2022a}
{Begley}, R., {Cullen}, F., {McLure}, R.~J., {et~al.} 2022, \mnras, 513, 3510,
  \dodoi{10.1093/mnras/stac1067}

\bibitem[{{Bian} {et~al.}(2017){Bian}, {Fan}, {McGreer}, {Cai}, \&
  {Jiang}}]{Bian2017a}
{Bian}, F., {Fan}, X., {McGreer}, I., {Cai}, Z., \& {Jiang}, L. 2017, \apjl,
  837, L12, \dodoi{10.3847/2041-8213/aa5ff7}

\bibitem[{{Bouwens} {et~al.}(2016){Bouwens}, {Smit}, {Labb{\'e}}, {Franx},
  {Caruana}, {Oesch}, {Stefanon}, \& {Rasappu}}]{Bouwens2016a}
{Bouwens}, R.~J., {Smit}, R., {Labb{\'e}}, I., {et~al.} 2016, \apj, 831, 176,
  \dodoi{10.3847/0004-637X/831/2/176}

\bibitem[{{Bouwens} {et~al.}(2012){Bouwens}, {Illingworth}, {Oesch}, {Trenti},
  {Labb{\'e}}, {Franx}, {Stiavelli}, {Carollo}, {van Dokkum}, \&
  {Magee}}]{Bouwens2012a}
{Bouwens}, R.~J., {Illingworth}, G.~D., {Oesch}, P.~A., {et~al.} 2012, \apjl,
  752, L5, \dodoi{10.1088/2041-8205/752/1/L5}

\bibitem[{Bradley {et~al.}(2023)Bradley, Sip{\H o}cz, Robitaille, Tollerud,
  Vin{\'{\i}}cius, Deil, Barbary, Wilson, Busko, Donath, G{\"u}nther, Cara,
  Lim, Me{\ss}linger, Conseil, Bostroem, Droettboom, Bray, Bratholm, Barentsen,
  Craig, Rathi, Pascual, Perren, Georgiev, de~Val-Borro, Kerzendorf, Bach,
  Quint, \& Souchereau}]{Bradley2023a}
Bradley, L., Sip{\H o}cz, B., Robitaille, T., {et~al.} 2023, astropy/photutils:
  1.8.0, 1.8.0,  Zenodo, \dodoi{10.5281/zenodo.7946442}

\bibitem[{{Bunker} {et~al.}(2023){Bunker}, {Saxena}, {Cameron}, {Willott},
  {Curtis-Lake}, {Jakobsen}, {Carniani}, {Smit}, {Maiolino}, {Witstok},
  {Curti}, {D'Eugenio}, {Jones}, {Ferruit}, {Arribas}, {Charlot}, {Chevallard},
  {Giardino}, {de Graaff}, {Looser}, {Luetzgendorf}, {Maseda}, {Rawle}, {Rix},
  {Rodriguez Del Pino}, {Alberts}, {Egami}, {Eisenstein}, {Endsley},
  {Hainline}, {Hausen}, {Johnson}, {Rieke}, {Rieke}, {Robertson}, {Shivaei},
  {Stark}, {Sun}, {Tacchella}, {Tang}, {Williams}, {Willmer}, {Baker}, {Baum},
  {Bhatawdekar}, {Bowler}, {Boyett}, {Chen}, {Circosta}, {Helton}, {Ji}, {Lyu},
  {Nelson}, {Parlanti}, {Perna}, {Sandles}, {Scholtz}, {Suess}, {Topping},
  {Uebler}, {Wallace}, \& {Whitler}}]{Bunker2023a}
{Bunker}, A.~J., {Saxena}, A., {Cameron}, A.~J., {et~al.} 2023, arXiv e-prints,
  arXiv:2302.07256, \dodoi{10.48550/arXiv.2302.07256}

\bibitem[{{Byler} {et~al.}(2017){Byler}, {Dalcanton}, {Conroy}, \&
  {Johnson}}]{Byler2017a}
{Byler}, N., {Dalcanton}, J.~J., {Conroy}, C., \& {Johnson}, B.~D. 2017, \apj,
  840, 44, \dodoi{10.3847/1538-4357/aa6c66}

\bibitem[{{Calzetti} {et~al.}(2000){Calzetti}, {Armus}, {Bohlin}, {Kinney},
  {Koornneef}, \& {Storchi-Bergmann}}]{Calzetti2000a}
{Calzetti}, D., {Armus}, L., {Bohlin}, R.~C., {et~al.} 2000, \apj, 533, 682,
  \dodoi{10.1086/308692}

\bibitem[{{Cameron} {et~al.}(2023){Cameron}, {Saxena}, {Bunker}, {D'Eugenio},
  {Carniani}, {Maiolino}, {Curtis-Lake}, {Ferruit}, {Jakobsen}, {Arribas},
  {Bonaventura}, {Charlot}, {Chevallard}, {Curti}, {Looser}, {Maseda}, {Rawle},
  {Rodr{\'\i}guez Del Pino}, {Smit}, {{\"U}bler}, {Willott}, {Witstok},
  {Egami}, {Eisenstein}, {Johnson}, {Hainline}, {Rieke}, {Robertson}, {Stark},
  {Tacchella}, {Williams}, {Bhatawdekar}, {Bowler}, {Boyett}, {Circosta},
  {Helton}, {Jones}, {Kumari}, {Ji}, {Nelson}, {Parlanti}, {Sandles},
  {Scholtz}, \& {Sun}}]{Cameron2023a}
{Cameron}, A.~J., {Saxena}, A., {Bunker}, A.~J., {et~al.} 2023, arXiv e-prints,
  arXiv:2302.04298, \dodoi{10.48550/arXiv.2302.04298}

\bibitem[{{Caminha} {et~al.}(2017){Caminha}, {Grillo}, {Rosati}, {Meneghetti},
  {Mercurio}, {Ettori}, {Balestra}, {Biviano}, {Umetsu}, {Vanzella},
  {Annunziatella}, {Bonamigo}, {Delgado-Correal}, {Girardi}, {Lombardi},
  {Nonino}, {Sartoris}, {Tozzi}, {Bartelmann}, {Bradley}, {Caputi}, {Coe},
  {Ford}, {Fritz}, {Gobat}, {Postman}, {Seitz}, \& {Zitrin}}]{Caminha2017a}
{Caminha}, G.~B., {Grillo}, C., {Rosati}, P., {et~al.} 2017, \aap, 607, A93,
  \dodoi{10.1051/0004-6361/201731498}

\bibitem[{{Caminha} {et~al.}(2019){Caminha}, {Rosati}, {Grillo}, {Rosani},
  {Caputi}, {Meneghetti}, {Mercurio}, {Balestra}, {Bergamini}, {Biviano},
  {Nonino}, {Umetsu}, {Vanzella}, {Annunziatella}, {Broadhurst},
  {Delgado-Correal}, {Demarco}, {Koekemoer}, {Lombardi}, {Maier}, {Verdugo}, \&
  {Zitrin}}]{Caminha2019a}
{Caminha}, G.~B., {Rosati}, P., {Grillo}, C., {et~al.} 2019, \aap, 632, A36,
  \dodoi{10.1051/0004-6361/201935454}

\bibitem[{{Cen} \& {Kimm}(2015)}]{Cen2015a}
{Cen}, R., \& {Kimm}, T. 2015, \apjl, 801, L25,
  \dodoi{10.1088/2041-8205/801/2/L25}

\bibitem[{{Chisholm} {et~al.}(2020){Chisholm}, {Prochaska}, {Schaerer},
  {Gazagnes}, \& {Henry}}]{Chisholm2020a}
{Chisholm}, J., {Prochaska}, J.~X., {Schaerer}, D., {Gazagnes}, S., \& {Henry},
  A. 2020, \mnras, 498, 2554, \dodoi{10.1093/mnras/staa2470}

\bibitem[{{Chisholm} {et~al.}(2018){Chisholm}, {Gazagnes}, {Schaerer},
  {Verhamme}, {Rigby}, {Bayliss}, {Sharon}, {Gladders}, \&
  {Dahle}}]{Chisholm2018a}
{Chisholm}, J., {Gazagnes}, S., {Schaerer}, D., {et~al.} 2018, \aap, 616, A30,
  \dodoi{10.1051/0004-6361/201832758}

\bibitem[{{Chisholm} {et~al.}(2022){Chisholm}, {Saldana-Lopez}, {Flury},
  {Schaerer}, {Jaskot}, {Amor{\'\i}n}, {Atek}, {Finkelstein}, {Fleming},
  {Ferguson}, {Fern{\'a}ndez}, {Giavalisco}, {Hayes}, {Heckman}, {Henry}, {Ji},
  {Marques-Chaves}, {Mauerhofer}, {McCandliss}, {Oey}, {{\"O}stlin},
  {Rutkowski}, {Scarlata}, {Thuan}, {Trebitsch}, {Wang}, {Worseck}, \&
  {Xu}}]{Chisholm2022a}
{Chisholm}, J., {Saldana-Lopez}, A., {Flury}, S., {et~al.} 2022, \mnras, 517,
  5104, \dodoi{10.1093/mnras/stac2874}

\bibitem[{{Choi} {et~al.}(2016){Choi}, {Dotter}, {Conroy}, {Cantiello},
  {Paxton}, \& {Johnson}}]{Choi2016a}
{Choi}, J., {Dotter}, A., {Conroy}, C., {et~al.} 2016, \apj, 823, 102,
  \dodoi{10.3847/0004-637X/823/2/102}

\bibitem[{{Conroy} {et~al.}(2009){Conroy}, {Gunn}, \& {White}}]{Conroy2009a}
{Conroy}, C., {Gunn}, J.~E., \& {White}, M. 2009, \apj, 699, 486,
  \dodoi{10.1088/0004-637X/699/1/486}

\bibitem[{{Curtis-Lake} {et~al.}(2022){Curtis-Lake}, {Carniani}, {Cameron},
  {Charlot}, {Jakobsen}, {Maiolino}, {Bunker}, {Witstok}, {Smit}, {Chevallard},
  {Willott}, {Ferruit}, {Arribas}, {Bonaventura}, {Curti}, {D'Eugenio},
  {Franx}, {Giardino}, {Looser}, {L{\"u}tzgendorf}, {Maseda}, {Rawle}, {Rix},
  {Rodriguez del Pino}, {{\"U}bler}, {Sirianni}, {Dressler}, {Egami},
  {Eisenstein}, {Endsley}, {Hainline}, {Hausen}, {Johnson}, {Rieke},
  {Robertson}, {Shivaei}, {Stark}, {Tacchella}, {Williams}, {Willmer},
  {Bhatawdekar}, {Bowler}, {Boyett}, {Chen}, {de Graaff}, {Helton}, {Hviding},
  {Jones}, {Kumari}, {Lyu}, {Nelson}, {Perna}, {Sandles}, {Saxena}, {Suess},
  {Sun}, {Topping}, {Wallace}, \& {Whitler}}]{Curtis-Lake2022a}
{Curtis-Lake}, E., {Carniani}, S., {Cameron}, A., {et~al.} 2022, arXiv
  e-prints, arXiv:2212.04568.
\newblock \doarXiv{2212.04568}

\bibitem[{{de Barros} {et~al.}(2016){de Barros}, {Vanzella}, {Amor{\'\i}n},
  {Castellano}, {Siana}, {Grazian}, {Suh}, {Balestra}, {Vignali}, {Verhamme},
  {Zamorani}, {Mignoli}, {Hasinger}, {Comastri}, {Pentericci},
  {P{\'e}rez-Montero}, {Fontana}, {Giavalisco}, \& {Gilli}}]{deBarros2016a}
{de Barros}, S., {Vanzella}, E., {Amor{\'\i}n}, R., {et~al.} 2016, \aap, 585,
  A51, \dodoi{10.1051/0004-6361/201527046}

\bibitem[{{Dijkstra} {et~al.}(2016){Dijkstra}, {Gronke}, \&
  {Venkatesan}}]{Dijkstra2016a}
{Dijkstra}, M., {Gronke}, M., \& {Venkatesan}, A. 2016, \apj, 828, 71,
  \dodoi{10.3847/0004-637X/828/2/71}

\bibitem[{{Dom{\'\i}nguez} {et~al.}(2015){Dom{\'\i}nguez}, {Siana}, {Brooks},
  {Christensen}, {Bruzual}, {Stark}, \& {Alavi}}]{Dominguez2015a}
{Dom{\'\i}nguez}, A., {Siana}, B., {Brooks}, A.~M., {et~al.} 2015, \mnras, 451,
  839, \dodoi{10.1093/mnras/stv1001}

\bibitem[{{Donnan} {et~al.}(2023){Donnan}, {McLeod}, {Dunlop}, {McLure},
  {Carnall}, {Begley}, {Cullen}, {Hamadouche}, {Bowler}, {Magee}, {McCracken},
  {Milvang-Jensen}, {Moneti}, \& {Targett}}]{Donnan2023a}
{Donnan}, C.~T., {McLeod}, D.~J., {Dunlop}, J.~S., {et~al.} 2023, \mnras, 518,
  6011, \dodoi{10.1093/mnras/stac3472}

\bibitem[{{Ferland} {et~al.}(2013){Ferland}, {Porter}, {van Hoof}, {Williams},
  {Abel}, {Lykins}, {Shaw}, {Henney}, \& {Stancil}}]{Ferland2013a}
{Ferland}, G.~J., {Porter}, R.~L., {van Hoof}, P.~A.~M., {et~al.} 2013, \rmxaa,
  49, 137, \dodoi{10.48550/arXiv.1302.4485}

\bibitem[{{Finkelstein} {et~al.}(2012){Finkelstein}, {Papovich}, {Salmon},
  {Finlator}, {Dickinson}, {Ferguson}, {Giavalisco}, {Koekemoer}, {Reddy},
  {Bassett}, {Conselice}, {Dunlop}, {Faber}, {Grogin}, {Hathi}, {Kocevski},
  {Lai}, {Lee}, {McLure}, {Mobasher}, \& {Newman}}]{Finkelstein2012a}
{Finkelstein}, S.~L., {Papovich}, C., {Salmon}, B., {et~al.} 2012, \apj, 756,
  164, \dodoi{10.1088/0004-637X/756/2/164}

\bibitem[{{Finkelstein} {et~al.}(2023){Finkelstein}, {Leung}, {Bagley},
  {Dickinson}, {Ferguson}, {Papovich}, {Akins}, {Arrabal Haro}, {Dave},
  {Dekel}, {Kartaltepe}, {Kocevski}, {Koekemoer}, {Pirzkal}, {Somerville},
  {Yung}, {Amorin}, {Backhaus}, {Behroozi}, {Bisigello}, {Bromm}, {Casey},
  {Chavez Ortiz}, {Cheng}, {Chworowsky}, {Cleri}, {Cooper}, {Davis}, {de la
  Vega}, {Elbaz}, {Franco}, {Fontana}, {Fujimoto}, {Giavalisco}, {Grogin},
  {Holwerda}, {Huertas-Company}, {Hirschmann}, {Iyer}, {Jogee}, {Jung},
  {Larson}, {Lucas}, {Mobasher}, {Morales}, {Morley}, {Mukherjee},
  {Perez-Gonzalez}, {Ravindranath}, {Rodighiero}, {Rowland}, {Tacchella},
  {Taylor}, {Trump}, \& {Wilkins}}]{Finkelstein2023b}
{Finkelstein}, S.~L., {Leung}, G. C.~K., {Bagley}, M.~B., {et~al.} 2023, arXiv
  e-prints, arXiv:2311.04279, \dodoi{10.48550/arXiv.2311.04279}

\bibitem[{{Fletcher} {et~al.}(2019){Fletcher}, {Tang}, {Robertson}, {Nakajima},
  {Ellis}, {Stark}, \& {Inoue}}]{Fletcher2019a}
{Fletcher}, T.~J., {Tang}, M., {Robertson}, B.~E., {et~al.} 2019, \apj, 878,
  87, \dodoi{10.3847/1538-4357/ab2045}

\bibitem[{{Flury} {et~al.}(2022{\natexlab{a}}){Flury}, {Jaskot}, {Ferguson},
  {Worseck}, {Makan}, {Chisholm}, {Saldana-Lopez}, {Schaerer}, {McCandliss},
  {Wang}, {Ford}, {Heckman}, {Ji}, {Giavalisco}, {Amorin}, {Atek}, {Blaizot},
  {Borthakur}, {Carr}, {Castellano}, {Cristiani}, {De Barros}, {Dickinson},
  {Finkelstein}, {Fleming}, {Fontanot}, {Garel}, {Grazian}, {Hayes}, {Henry},
  {Mauerhofer}, {Micheva}, {Oey}, {Ostlin}, {Papovich}, {Pentericci},
  {Ravindranath}, {Rosdahl}, {Rutkowski}, {Santini}, {Scarlata}, {Teplitz},
  {Thuan}, {Trebitsch}, {Vanzella}, {Verhamme}, \& {Xu}}]{Flury2022a}
{Flury}, S.~R., {Jaskot}, A.~E., {Ferguson}, H.~C., {et~al.}
  2022{\natexlab{a}}, \apjs, 260, 1, \dodoi{10.3847/1538-4365/ac5331}

\bibitem[{{Flury} {et~al.}(2022{\natexlab{b}}){Flury}, {Jaskot}, {Ferguson},
  {Worseck}, {Makan}, {Chisholm}, {Saldana-Lopez}, {Schaerer}, {McCandliss},
  {Xu}, {Wang}, {Oey}, {Ford}, {Heckman}, {Ji}, {Giavalisco}, {Amor{\'\i}n},
  {Atek}, {Blaizot}, {Borthakur}, {Carr}, {Castellano}, {Barros}, {Dickinson},
  {Finkelstein}, {Fleming}, {Fontanot}, {Garel}, {Grazian}, {Hayes}, {Henry},
  {Mauerhofer}, {Micheva}, {Ostlin}, {Papovich}, {Pentericci}, {Ravindranath},
  {Rosdahl}, {Rutkowski}, {Santini}, {Scarlata}, {Teplitz}, {Thuan},
  {Trebitsch}, {Vanzella}, \& {Verhamme}}]{Flury2022b}
---. 2022{\natexlab{b}}, \apj, 930, 126, \dodoi{10.3847/1538-4357/ac61e4}

\bibitem[{{Fruchter} \& {et al.}(2010)}]{Fruchter2010a}
{Fruchter}, A.~S., \& {et al.} 2010, in 2010 Space Telescope Science Institute
  Calibration Workshop, 382--387

\bibitem[{{Fruchter} \& {Hook}(2002)}]{Fruchter2002a}
{Fruchter}, A.~S., \& {Hook}, R.~N. 2002, \pasp, 114, 144,
  \dodoi{10.1086/338393}

\bibitem[{{Fujimoto} {et~al.}(2023){Fujimoto}, {Arrabal Haro}, {Dickinson},
  {Finkelstein}, {Kartaltepe}, {Larson}, {Burgarella}, {Bagley}, {Behroozi},
  {Chworowsky}, {Hirschmann}, {Trump}, {Wilkins}, {Yung}, {Koekemoer},
  {Papovich}, {Pirzkal}, {Ferguson}, {Fontana}, {Grogin}, {Grazian}, {Kewley},
  {Kocevski}, {Lotz}, {Pentericci}, {Ravindranath}, {Somerville}, {Amorin},
  {Backhaus}, {Calabro}, {Casey}, {Cooper}, {Franco}, {Giavalisco}, {Hathi},
  {Harish}, {Hutchison}, {Iyer}, {Jung}, {Lucas}, \& {Zavala}}]{Fujimoto2023a}
{Fujimoto}, S., {Arrabal Haro}, P., {Dickinson}, M., {et~al.} 2023, arXiv
  e-prints, arXiv:2301.09482.
\newblock \doarXiv{2301.09482}

\bibitem[{{Grazian} {et~al.}(2017){Grazian}, {Giallongo}, {Paris}, {Boutsia},
  {Dickinson}, {Santini}, {Windhorst}, {Jansen}, {Cohen}, {Ashcraft},
  {Scarlata}, {Rutkowski}, {Vanzella}, {Cusano}, {Cristiani}, {Giavalisco},
  {Ferguson}, {Koekemoer}, {Grogin}, {Castellano}, {Fiore}, {Fontana},
  {Marchi}, {Pedichini}, {Pentericci}, {Amor{\'\i}n}, {Barro}, {Bonchi},
  {Bongiorno}, {Faber}, {Fumana}, {Galametz}, {Guaita}, {Kocevski}, {Merlin},
  {Nonino}, {O'Connell}, {Pilo}, {Ryan}, {Sani}, {Speziali}, {Testa}, {Weiner},
  \& {Yan}}]{Grazian2017a}
{Grazian}, A., {Giallongo}, E., {Paris}, D., {et~al.} 2017, \aap, 602, A18,
  \dodoi{10.1051/0004-6361/201730447}

\bibitem[{{Harikane} {et~al.}(2023){Harikane}, {Ouchi}, {Oguri}, {Ono},
  {Nakajima}, {Isobe}, {Umeda}, {Mawatari}, \& {Zhang}}]{Harikane2023a}
{Harikane}, Y., {Ouchi}, M., {Oguri}, M., {et~al.} 2023, \apjs, 265, 5,
  \dodoi{10.3847/1538-4365/acaaa9}

\bibitem[{Harris {et~al.}(2020)Harris, Millman, van~der Walt, Gommers,
  Virtanen, Cournapeau, Wieser, Taylor, Berg, Smith, Kern, Picus, Hoyer, van
  Kerkwijk, Brett, Haldane, del R{\'{i}}o, Wiebe, Peterson,
  G{\'{e}}rard-Marchant, Sheppard, Reddy, Weckesser, Abbasi, Gohlke, \&
  Oliphant}]{Harris2020a}
Harris, C.~R., Millman, K.~J., van~der Walt, S.~J., {et~al.} 2020, Nature, 585,
  357, \dodoi{10.1038/s41586-020-2649-2}

\bibitem[{{Hayes}(2015)}]{Hayes2015a}
{Hayes}, M. 2015, \pasa, 32, e027, \dodoi{10.1017/pasa.2015.25}

\bibitem[{{Henry} {et~al.}(2018){Henry}, {Berg}, {Scarlata}, {Verhamme}, \&
  {Erb}}]{Henry2018a}
{Henry}, A., {Berg}, D.~A., {Scarlata}, C., {Verhamme}, A., \& {Erb}, D. 2018,
  \apj, 855, 96, \dodoi{10.3847/1538-4357/aab099}

\bibitem[{{Hu} {et~al.}(2023){Hu}, {Martin}, {Gronke}, {Gazagnes}, {Hayes},
  {Chisholm}, {Heckman}, {Mingozzi}, {Roy}, {Senchyna}, {Xu}, {Berg}, {James},
  {Stark}, {Arellano-C{\'o}rdova}, {Henry}, {Jaskot}, {Kumari}, {Parker},
  {Scarlata}, {Wofford}, {Amor{\'\i}n}, {Leonhardes-Barboza}, {Brinchmann},
  {Carr}, \& {Aloisi}}]{Hu2023a}
{Hu}, W., {Martin}, C.~L., {Gronke}, M., {et~al.} 2023, \apj, 956, 39,
  \dodoi{10.3847/1538-4357/aceefd}

\bibitem[{Hunter(2007)}]{Hunter2007a}
Hunter, J.~D. 2007, Computing in Science \& Engineering, 9, 90,
  \dodoi{10.1109/MCSE.2007.55}

\bibitem[{{Inoue} {et~al.}(2014){Inoue}, {Shimizu}, {Iwata}, \&
  {Tanaka}}]{Inoue2014a}
{Inoue}, A.~K., {Shimizu}, I., {Iwata}, I., \& {Tanaka}, M. 2014, \mnras, 442,
  1805, \dodoi{10.1093/mnras/stu936}

\bibitem[{{Iwata} {et~al.}(2009){Iwata}, {Inoue}, {Matsuda}, {Furusawa},
  {Hayashino}, {Kousai}, {Akiyama}, {Yamada}, {Burgarella}, \&
  {Deharveng}}]{Iwata2009a}
{Iwata}, I., {Inoue}, A.~K., {Matsuda}, Y., {et~al.} 2009, \apj, 692, 1287,
  \dodoi{10.1088/0004-637X/692/2/1287}

\bibitem[{{Izotov} {et~al.}(2016{\natexlab{a}}){Izotov}, {Orlitov{\'a}},
  {Schaerer}, {Thuan}, {Verhamme}, {Guseva}, \& {Worseck}}]{Izotov2016a}
{Izotov}, Y.~I., {Orlitov{\'a}}, I., {Schaerer}, D., {et~al.}
  2016{\natexlab{a}}, \nat, 529, 178, \dodoi{10.1038/nature16456}

\bibitem[{{Izotov} {et~al.}(2016{\natexlab{b}}){Izotov}, {Schaerer}, {Thuan},
  {Worseck}, {Guseva}, {Orlitov{\'a}}, \& {Verhamme}}]{Izotov2016b}
{Izotov}, Y.~I., {Schaerer}, D., {Thuan}, T.~X., {et~al.} 2016{\natexlab{b}},
  \mnras, 461, 3683, \dodoi{10.1093/mnras/stw1205}

\bibitem[{{Izotov} {et~al.}(2018{\natexlab{a}}){Izotov}, {Schaerer}, {Worseck},
  {Guseva}, {Thuan}, {Verhamme}, {Orlitov{\'a}}, \& {Fricke}}]{Izotov2018a}
{Izotov}, Y.~I., {Schaerer}, D., {Worseck}, G., {et~al.} 2018{\natexlab{a}},
  \mnras, 474, 4514, \dodoi{10.1093/mnras/stx3115}

\bibitem[{{Izotov} {et~al.}(2020){Izotov}, {Schaerer}, {Worseck}, {Verhamme},
  {Guseva}, {Thuan}, {Orlitov{\'a}}, \& {Fricke}}]{Izotov2020a}
---. 2020, \mnras, 491, 468, \dodoi{10.1093/mnras/stz3041}

\bibitem[{{Izotov} {et~al.}(2021){Izotov}, {Worseck}, {Schaerer}, {Guseva},
  {Chisholm}, {Thuan}, {Fricke}, \& {Verhamme}}]{Izotov2021a}
{Izotov}, Y.~I., {Worseck}, G., {Schaerer}, D., {et~al.} 2021, \mnras, 503,
  1734, \dodoi{10.1093/mnras/stab612}

\bibitem[{{Izotov} {et~al.}(2018{\natexlab{b}}){Izotov}, {Worseck}, {Schaerer},
  {Guseva}, {Thuan}, {Fricke}, \& {Orlitov{\'a}}}]{Izotov2018b}
---. 2018{\natexlab{b}}, \mnras, 478, 4851, \dodoi{10.1093/mnras/sty1378}

\bibitem[{{James} {et~al.}(2014){James}, {Aloisi}, {Heckman}, {Sohn}, \&
  {Wolfe}}]{James2014a}
{James}, B.~L., {Aloisi}, A., {Heckman}, T., {Sohn}, S.~T., \& {Wolfe}, M.~A.
  2014, \apj, 795, 109, \dodoi{10.1088/0004-637X/795/2/109}

\bibitem[{{Ji} {et~al.}(2020){Ji}, {Giavalisco}, {Vanzella}, {Siana},
  {Pentericci}, {Jaskot}, {Liu}, {Nonino}, {Ferguson}, {Castellano},
  {Mannucci}, {Schaerer}, {Fynbo}, {Papovich}, {Carnall}, {Amorin}, {Simons},
  {Hathi}, {Cullen}, \& {McLeod}}]{Ji2020a}
{Ji}, Z., {Giavalisco}, M., {Vanzella}, E., {et~al.} 2020, \apj, 888, 109,
  \dodoi{10.3847/1538-4357/ab5fdc}

\bibitem[{{Johnson} {et~al.}(2021){Johnson}, {Leja}, {Conroy}, \&
  {Speagle}}]{Johnson2021a}
{Johnson}, B.~D., {Leja}, J., {Conroy}, C., \& {Speagle}, J.~S. 2021, \apjs,
  254, 22, \dodoi{10.3847/1538-4365/abef67}

\bibitem[{{Jung} {et~al.}(2023){Jung}, {Finkelstein}, {Arrabal Haro},
  {Dickinson}, {Ferguson}, {Hutchison}, {Kartaltepe}, {Larson}, {Simons},
  {Papovich}, {Park}, {Pentericci}, {Trump}, {Amorin}, {Backhaus}, {Casey},
  {Cheng}, {Cleri}, {Cooper}, {Cooper}, {Gardner}, {Gawiser}, {Grazian},
  {Hathi}, {Hirschmann}, {Koekemoer}, {Lucas}, {Mobasher}, {Ravindranath},
  {Straughn}, {Yung}, \& {de la Vega}}]{Jung2023a}
{Jung}, I., {Finkelstein}, S.~L., {Arrabal Haro}, P., {et~al.} 2023, arXiv
  e-prints, arXiv:2304.05385, \dodoi{10.48550/arXiv.2304.05385}

\bibitem[{{Kerutt} {et~al.}(2023){Kerutt}, {Oesch}, {Wisotzki}, {Verhamme},
  {Atek}, {Herenz}, {Illingworth}, {Kusakabe}, {Matthee}, {Mauerhofer},
  {Montes}, {Naidu}, {Nelson}, {Reddy}, {Schaye}, {Simmonds}, {Urrutia}, \&
  {Vitte}}]{Kerutt2023a}
{Kerutt}, J., {Oesch}, P.~A., {Wisotzki}, L., {et~al.} 2023, arXiv e-prints,
  arXiv:2312.08791, \dodoi{10.48550/arXiv.2312.08791}

\bibitem[{{Kim} {et~al.}(2023){Kim}, {Bayliss}, {Rigby}, {Gladders},
  {Chisholm}, {Sharon}, {Dahle}, {Rivera-Thorsen}, {Florian}, {Khullar},
  {Mahler}, {Mainali}, {Napier}, {Navarre}, {Owens}, \& {Roberson}}]{Kim2023a}
{Kim}, K.~J., {Bayliss}, M.~B., {Rigby}, J.~R., {et~al.} 2023, arXiv e-prints,
  arXiv:2305.13405, \dodoi{10.48550/arXiv.2305.13405}

\bibitem[{{Kunth} {et~al.}(1994){Kunth}, {Lequeux}, {Sargent}, \&
  {Viallefond}}]{Kunth1994a}
{Kunth}, D., {Lequeux}, J., {Sargent}, W.~L.~W., \& {Viallefond}, F. 1994,
  \aap, 282, 709

\bibitem[{{Lam} {et~al.}(2019){Lam}, {Bouwens}, {Labb{\'e}}, {Schaye},
  {Schmidt}, {Maseda}, {Bacon}, {Boogaard}, {Nanayakkara}, {Richard}, {Mahler},
  \& {Urrutia}}]{Lam2019a}
{Lam}, D., {Bouwens}, R.~J., {Labb{\'e}}, I., {et~al.} 2019, \aap, 627, A164,
  \dodoi{10.1051/0004-6361/201935227}

\bibitem[{{Landsman}(1993)}]{Landsman1993a}
{Landsman}, W.~B. 1993, in Astronomical Society of the Pacific Conference
  Series, Vol.~52, Astronomical Data Analysis Software and Systems II, ed.
  R.~J. {Hanisch}, R.~J.~V. {Brissenden}, \& J.~{Barnes}, 246

\bibitem[{{Larson} {et~al.}(2023{\natexlab{a}}){Larson}, {Finkelstein},
  {Kocevski}, {Hutchison}, {Trump}, {Arrabal Haro}, {Bromm}, {Cleri},
  {Dickinson}, {Fujimoto}, {Kartaltepe}, {Koekemoer}, {Papovich}, {Pirzkal},
  {Tacchella}, {Zavala}, {Bagley}, {Behroozi}, {Champagne}, {Cole}, {Jung},
  {Morales}, {Yang}, {Zhang}, {Zitrin}, {Amor{\'\i}n}, {Burgarella}, {Casey},
  {Ch{\'a}vez Ortiz}, {Cox}, {Chworowsky}, {Fontana}, {Gawiser}, {Grazian},
  {Grogin}, {Harish}, {Hathi}, {Hirschmann}, {Holwerda}, {Juneau}, {Leung},
  {Lucas}, {McGrath}, {P{\'e}rez-Gonz{\'a}lez}, {Rigby}, {Seill{\'e}},
  {Simons}, {Weiner}, {Wilkins}, {Yung}, \& {The CEERS Team}}]{Larson2023a}
{Larson}, R.~L., {Finkelstein}, S.~L., {Kocevski}, D.~D., {et~al.}
  2023{\natexlab{a}}, arXiv e-prints, arXiv:2303.08918,
  \dodoi{10.48550/arXiv.2303.08918}

\bibitem[{{Larson} {et~al.}(2023{\natexlab{b}}){Larson}, {Hutchison}, {Bagley},
  {Finkelstein}, {Yung}, {Somerville}, {Hirschmann}, {Brammer}, {Holwerda},
  {Papovich}, {Morales}, \& {Wilkins}}]{Larson2023b}
{Larson}, R.~L., {Hutchison}, T.~A., {Bagley}, M., {et~al.} 2023{\natexlab{b}},
  \apj, 958, 141, \dodoi{10.3847/1538-4357/acfed4}

\bibitem[{{Leja} {et~al.}(2017){Leja}, {Johnson}, {Conroy}, {van Dokkum}, \&
  {Byler}}]{Leja2017a}
{Leja}, J., {Johnson}, B.~D., {Conroy}, C., {van Dokkum}, P.~G., \& {Byler}, N.
  2017, \apj, 837, 170, \dodoi{10.3847/1538-4357/aa5ffe}

\bibitem[{{Ma} {et~al.}(2015){Ma}, {Kasen}, {Hopkins}, {Faucher-Gigu{\`e}re},
  {Quataert}, {Kere{\v{s}}}, \& {Murray}}]{Ma2015a}
{Ma}, X., {Kasen}, D., {Hopkins}, P.~F., {et~al.} 2015, \mnras, 453, 960,
  \dodoi{10.1093/mnras/stv1679}

\bibitem[{{Ma} {et~al.}(2020){Ma}, {Quataert}, {Wetzel}, {Hopkins},
  {Faucher-Gigu{\`e}re}, \& {Kere{\v{s}}}}]{Ma2020a}
{Ma}, X., {Quataert}, E., {Wetzel}, A., {et~al.} 2020, \mnras, 498, 2001,
  \dodoi{10.1093/mnras/staa2404}

\bibitem[{{Madau}(1995)}]{Madau1995a}
{Madau}, P. 1995, \apj, 441, 18, \dodoi{10.1086/175332}

\bibitem[{{Madau} {et~al.}(1999){Madau}, {Haardt}, \& {Rees}}]{Madau1999a}
{Madau}, P., {Haardt}, F., \& {Rees}, M.~J. 1999, \apj, 514, 648,
  \dodoi{10.1086/306975}

\bibitem[{{Mainali} {et~al.}(2022){Mainali}, {Rigby}, {Chisholm}, {Bayliss},
  {Bordoloi}, {Gladders}, {Rivera-Thorsen}, {Dahle}, {Sharon}, {Florian},
  {Berg}, {Sharma}, {Owens}, {Kjellgren}, {Kim}, \& {Wayne}}]{Mainali2022a}
{Mainali}, R., {Rigby}, J.~R., {Chisholm}, J., {et~al.} 2022, \apj, 940, 160,
  \dodoi{10.3847/1538-4357/ac9cd6}

\bibitem[{{Maiolino} {et~al.}(2023){Maiolino}, {Scholtz}, {Witstok},
  {Carniani}, {D'Eugenio}, {de Graaff}, {Uebler}, {Tacchella}, {Curtis-Lake},
  {Arribas}, {Bunker}, {Charlot}, {Chevallard}, {Curti}, {Looser}, {Maseda},
  {Rawle}, {Rodriguez Del Pino}, {Willott}, {Egami}, {Eisenstein}, {Hainline},
  {Robertson}, {Williams}, {Willmer}, {Baker}, {Boyett}, {DeCoursey}, {Fabian},
  {Helton}, {Ji}, {Jones}, {Kumari}, {Laporte}, {Nelson}, {Perna}, {Sandles},
  {Shivaei}, \& {Sun}}]{Maiolino2023a}
{Maiolino}, R., {Scholtz}, J., {Witstok}, J., {et~al.} 2023, arXiv e-prints,
  arXiv:2305.12492, \dodoi{10.48550/arXiv.2305.12492}

\bibitem[{{Marques-Chaves} {et~al.}(2021){Marques-Chaves}, {Schaerer},
  {{\'A}lvarez-M{\'a}rquez}, {Colina}, {Dessauges-Zavadsky},
  {P{\'e}rez-Fournon}, {Saldana-Lopez}, \& {Verhamme}}]{Marques-Chaves2021a}
{Marques-Chaves}, R., {Schaerer}, D., {{\'A}lvarez-M{\'a}rquez}, J., {et~al.}
  2021, \mnras, 507, 524, \dodoi{10.1093/mnras/stab2187}

\bibitem[{{Mascia} {et~al.}(2023{\natexlab{a}}){Mascia}, {Pentericci},
  {Calabro'}, {Treu}, {Santini}, {Yang}, {Napolitano}, {Roberts-Borsani},
  {Bergamini}, {Grillo}, {Rosati}, {Vulcani}, {Castellano}, {Boyett},
  {Fontana}, {Glazebrook}, {Henry}, {Mason}, {Merlin}, {Morishita},
  {Nanayakkara}, {Paris}, {Roy}, {Williams}, {Wang}, {Brammer}, {Bradac},
  {Chen}, {Kelly}, {Koekemoer}, {Trenti}, \& {Windhorst}}]{Mascia2023a}
{Mascia}, S., {Pentericci}, L., {Calabro'}, A., {et~al.} 2023{\natexlab{a}},
  arXiv e-prints, arXiv:2301.02816, \dodoi{10.48550/arXiv.2301.02816}

\bibitem[{{Mascia} {et~al.}(2023{\natexlab{b}}){Mascia}, {Pentericci},
  {Calabr{\`o}}, {Santini}, {Napolitano}, {Arrabal Haro}, {Castellano},
  {Dickinson}, {Ocvirk}, {Lewis}, {Amor{\'\i}n}, {Bagley}, {Cleri},
  {Costantin}, {Dekel}, {Finkelstein}, {Fontana}, {Giavalisco}, {Grogin},
  {Hathi}, {Hirschmann}, {Holwerda}, {Jung}, {Kartaltepe}, {Koekemoer},
  {Lucas}, {Papovich}, {P{\'e}rez-Gonz{\'a}lez}, {Pirzkal}, {Trump}, {Wilkins},
  \& {Yung}}]{Mascia2023b}
{Mascia}, S., {Pentericci}, L., {Calabr{\`o}}, A., {et~al.} 2023{\natexlab{b}},
  arXiv e-prints, arXiv:2309.02219, \dodoi{10.48550/arXiv.2309.02219}

\bibitem[{{Matthee} {et~al.}(2017){Matthee}, {Sobral}, {Best}, {Khostovan},
  {Oteo}, {Bouwens}, \& {R{\"o}ttgering}}]{Matthee2017a}
{Matthee}, J., {Sobral}, D., {Best}, P., {et~al.} 2017, \mnras, 465, 3637,
  \dodoi{10.1093/mnras/stw2973}

\bibitem[{{McLeod} {et~al.}(2023){McLeod}, {Donnan}, {McLure}, {Dunlop},
  {Magee}, {Begley}, {Carnall}, {Cullen}, {Ellis}, {Hamadouche}, \&
  {Stanton}}]{McLeod2023a}
{McLeod}, D.~J., {Donnan}, C.~T., {McLure}, R.~J., {et~al.} 2023, \mnras,
  \dodoi{10.1093/mnras/stad3471}

\bibitem[{{Morales} {et~al.}(2023){Morales}, {Finkelstein}, {Leung}, {Bagley},
  {Cleri}, {Dave}, {Dickinson}, {Ferguson}, {Hathi}, {Jones}, {Koekemoer},
  {Papovich}, {Perez-Gonzalez}, {Pirzkal}, {Smith}, {Wilkins}, \&
  {Yung}}]{Morales2023a}
{Morales}, A.~M., {Finkelstein}, S.~L., {Leung}, G. C.~K., {et~al.} 2023, arXiv
  e-prints, arXiv:2311.04294, \dodoi{10.48550/arXiv.2311.04294}

\bibitem[{{Mostardi} {et~al.}(2015){Mostardi}, {Shapley}, {Steidel}, {Trainor},
  {Reddy}, \& {Siana}}]{Mostardi2015a}
{Mostardi}, R.~E., {Shapley}, A.~E., {Steidel}, C.~C., {et~al.} 2015, \apj,
  810, 107, \dodoi{10.1088/0004-637X/810/2/107}

\bibitem[{{Nakajima} {et~al.}(2016){Nakajima}, {Ellis}, {Iwata}, {Inoue},
  {Kusakabe}, {Ouchi}, \& {Robertson}}]{Nakajima2016a}
{Nakajima}, K., {Ellis}, R.~S., {Iwata}, I., {et~al.} 2016, \apjl, 831, L9,
  \dodoi{10.3847/2041-8205/831/1/L9}

\bibitem[{{Oke} \& {Gunn}(1983)}]{Oke1983a}
{Oke}, J.~B., \& {Gunn}, J.~E. 1983, \apj, 266, 713, \dodoi{10.1086/160817}

\bibitem[{{Paardekooper} {et~al.}(2015){Paardekooper}, {Khochfar}, \& {Dalla
  Vecchia}}]{Paardekooper2015a}
{Paardekooper}, J.-P., {Khochfar}, S., \& {Dalla Vecchia}, C. 2015, \mnras,
  451, 2544, \dodoi{10.1093/mnras/stv1114}

\bibitem[{{Pahl} {et~al.}(2021){Pahl}, {Shapley}, {Steidel}, {Chen}, \&
  {Reddy}}]{Pahl2021a}
{Pahl}, A.~J., {Shapley}, A., {Steidel}, C.~C., {Chen}, Y., \& {Reddy}, N.~A.
  2021, \mnras, 505, 2447, \dodoi{10.1093/mnras/stab1374}

\bibitem[{{Pei}(1992)}]{Pei1992a}
{Pei}, Y.~C. 1992, \apj, 395, 130, \dodoi{10.1086/171637}

\bibitem[{{Planck Collaboration} {et~al.}(2016){Planck Collaboration}, {Ade},
  {Aghanim}, {Arnaud}, {Ashdown}, {Aumont}, {Baccigalupi}, {Banday},
  {Barreiro}, {Bartlett}, \& et~al.}]{Planck-Collaboration2016a}
{Planck Collaboration}, {Ade}, P.~A.~R., {Aghanim}, N., {et~al.} 2016, \aap,
  594, A13, \dodoi{10.1051/0004-6361/201525830}

\bibitem[{{Postman} {et~al.}(2012){Postman}, {Coe}, {Ben{\'\i}tez}, {Bradley},
  {Broadhurst}, {Donahue}, {Ford}, {Graur}, {Graves}, {Jouvel}, {Koekemoer},
  {Lemze}, {Medezinski}, {Molino}, {Moustakas}, {Ogaz}, {Riess}, {Rodney},
  {Rosati}, {Umetsu}, {Zheng}, {Zitrin}, {Bartelmann}, {Bouwens}, {Czakon},
  {Golwala}, {Host}, {Infante}, {Jha}, {Jimenez-Teja}, {Kelson}, {Lahav},
  {Lazkoz}, {Maoz}, {McCully}, {Melchior}, {Meneghetti}, {Merten}, {Moustakas},
  {Nonino}, {Patel}, {Reg{\"o}s}, {Sayers}, {Seitz}, \& {Van der
  Wel}}]{Postman2012a}
{Postman}, M., {Coe}, D., {Ben{\'\i}tez}, N., {et~al.} 2012, \apjs, 199, 25,
  \dodoi{10.1088/0067-0049/199/2/25}

\bibitem[{{Richard} {et~al.}(2021){Richard}, {Claeyssens}, {Lagattuta},
  {Guaita}, {Bauer}, {Pello}, {Carton}, {Bacon}, {Soucail}, {Lyon}, {Kneib},
  {Mahler}, {Cl{\'e}ment}, {Mercier}, {Variu}, {Tamone}, {Ebeling}, {Schmidt},
  {Nanayakkara}, {Maseda}, {Weilbacher}, {Bouch{\'e}}, {Bouwens}, {Wisotzki},
  {de la Vieuville}, {Martinez}, \& {Patr{\'\i}cio}}]{Richard2021a}
{Richard}, J., {Claeyssens}, A., {Lagattuta}, D., {et~al.} 2021, \aap, 646,
  A83, \dodoi{10.1051/0004-6361/202039462}

\bibitem[{{Rivera-Thorsen} {et~al.}(2017){Rivera-Thorsen}, {Dahle}, {Gronke},
  {Bayliss}, {Rigby}, {Simcoe}, {Bordoloi}, {Turner}, \&
  {Furesz}}]{Rivera-Thorsen2017a}
{Rivera-Thorsen}, T.~E., {Dahle}, H., {Gronke}, M., {et~al.} 2017, \aap, 608,
  L4, \dodoi{10.1051/0004-6361/201732173}

\bibitem[{{Rivera-Thorsen} {et~al.}(2019){Rivera-Thorsen}, {Dahle}, {Chisholm},
  {Florian}, {Gronke}, {Rigby}, {Gladders}, {Mahler}, {Sharon}, \&
  {Bayliss}}]{Rivera-Thorsen2019a}
{Rivera-Thorsen}, T.~E., {Dahle}, H., {Chisholm}, J., {et~al.} 2019, Science,
  366, 738, \dodoi{10.1126/science.aaw0978}

\bibitem[{{Roberts-Borsani} {et~al.}(2022){Roberts-Borsani}, {Treu}, {Mason},
  {Ellis}, {Laporte}, {Schmidt}, {Brada{\v{c}}}, {Fontana}, {Morishita}, \&
  {Santini}}]{Roberts-Borsani2022a}
{Roberts-Borsani}, G., {Treu}, T., {Mason}, C., {et~al.} 2022, arXiv e-prints,
  arXiv:2207.01629.
\newblock \doarXiv{2207.01629}

\bibitem[{{Robertson}(2022)}]{Robertson2022a}
{Robertson}, B.~E. 2022, \araa, 60, 121,
  \dodoi{10.1146/annurev-astro-120221-044656}

\bibitem[{{Rosdahl} {et~al.}(2022){Rosdahl}, {Blaizot}, {Katz}, {Kimm},
  {Garel}, {Haehnelt}, {Keating}, {Martin-Alvarez}, {Michel-Dansac}, \&
  {Ocvirk}}]{Rosdahl2022a}
{Rosdahl}, J., {Blaizot}, J., {Katz}, H., {et~al.} 2022, \mnras, 515, 2386,
  \dodoi{10.1093/mnras/stac194210.48550/arXiv.2207.03232}

\bibitem[{{Runnholm} {et~al.}(2023){Runnholm}, {Hayes}, {Lin}, {Melinder},
  {Scarlata}, {Adamo}, {Augustin}, {Bik}, {Blaizot}, {Cannon}, {Cantalupo},
  {Garel}, {Gronke}, {Herenz}, {Leclercq}, {{\"O}stlin}, {Peroux}, {Rasekh},
  {Rutkowski}, {Verhamme}, \& {Wisotzki}}]{Runnholm2023a}
{Runnholm}, A., {Hayes}, M.~J., {Lin}, Y.-H., {et~al.} 2023, \mnras, 522, 4275,
  \dodoi{10.1093/mnras/stad1264}

\bibitem[{{Rutkowski} {et~al.}(2016){Rutkowski}, {Scarlata}, {Haardt}, {Siana},
  {Henry}, {Rafelski}, {Hayes}, {Salvato}, {Pahl}, {Mehta}, {Beck}, {Malkan},
  \& {Teplitz}}]{Rutkowski2016a}
{Rutkowski}, M.~J., {Scarlata}, C., {Haardt}, F., {et~al.} 2016, \apj, 819, 81,
  \dodoi{10.3847/0004-637X/819/1/81}

\bibitem[{{Rutkowski} {et~al.}(2017){Rutkowski}, {Scarlata}, {Henry}, {Hayes},
  {Mehta}, {Hathi}, {Cohen}, {Windhorst}, {Koekemoer}, {Teplitz}, {Haardt}, \&
  {Siana}}]{Rutkowski2017a}
{Rutkowski}, M.~J., {Scarlata}, C., {Henry}, A., {et~al.} 2017, \apjl, 841,
  L27, \dodoi{10.3847/2041-8213/aa733b}

\bibitem[{{Saha} {et~al.}(2020){Saha}, {Tandon}, {Simmonds}, {Verhamme},
  {Paswan}, {Schaerer}, {Rutkowski}, {Borgohain}, {Elmegreen}, {Inoue},
  {Combes}, {Elmegreen}, \& {Paalvast}}]{Saha2020a}
{Saha}, K., {Tandon}, S.~N., {Simmonds}, C., {et~al.} 2020, Nature Astronomy,
  4, 1185, \dodoi{10.1038/s41550-020-1173-5}

\bibitem[{{Saldana-Lopez} {et~al.}(2022){Saldana-Lopez}, {Schaerer},
  {Chisholm}, {Flury}, {Jaskot}, {Worseck}, {Makan}, {Gazagnes}, {Mauerhofer},
  {Verhamme}, {Amor{\'\i}n}, {Ferguson}, {Giavalisco}, {Grazian}, {Hayes},
  {Heckman}, {Henry}, {Ji}, {Marques-Chaves}, {McCandliss}, {Oey},
  {{\"O}stlin}, {Pentericci}, {Thuan}, {Trebitsch}, {Vanzella}, \&
  {Xu}}]{Saldana-Lopez2022a}
{Saldana-Lopez}, A., {Schaerer}, D., {Chisholm}, J., {et~al.} 2022, \aap, 663,
  A59, \dodoi{10.1051/0004-6361/202141864}

\bibitem[{{Saxena} {et~al.}(2023){Saxena}, {Bunker}, {Jones}, {Stark},
  {Cameron}, {Witstok}, {Arribas}, {Baker}, {Baum}, {Bhatawdekar}, {Bowler},
  {Boyett}, {Carniani}, {Charlot}, {Chevallard}, {Curti}, {Curtis-Lake},
  {Eisenstein}, {Endsley}, {Hainline}, {Helton}, {Johnson}, {Kumari}, {Looser},
  {Maiolino}, {Rieke}, {Rix}, {Robertson}, {Sandles}, {Simmonds}, {Smit},
  {Tacchella}, {Williams}, {Willmer}, \& {Willott}}]{Saxena2023b}
{Saxena}, A., {Bunker}, A.~J., {Jones}, G.~C., {et~al.} 2023, arXiv e-prints,
  arXiv:2306.04536, \dodoi{10.48550/arXiv.2306.04536}

\bibitem[{{Schaerer}(2003)}]{Schaerer2003a}
{Schaerer}, D. 2003, \aap, 397, 527, \dodoi{10.1051/0004-6361:20021525}

\bibitem[{{Schlegel} {et~al.}(1998){Schlegel}, {Finkbeiner}, \&
  {Davis}}]{Schlegel1998a}
{Schlegel}, D.~J., {Finkbeiner}, D.~P., \& {Davis}, M. 1998, \apj, 500, 525,
  \dodoi{10.1086/305772}

\bibitem[{{Shapley} {et~al.}(2006){Shapley}, {Steidel}, {Pettini},
  {Adelberger}, \& {Erb}}]{Shapley2006a}
{Shapley}, A.~E., {Steidel}, C.~C., {Pettini}, M., {Adelberger}, K.~L., \&
  {Erb}, D.~K. 2006, \apj, 651, 688, \dodoi{10.1086/507511}

\bibitem[{{Shapley} {et~al.}(2016){Shapley}, {Steidel}, {Strom},
  {Bogosavljevi{\'c}}, {Reddy}, {Siana}, {Mostardi}, \& {Rudie}}]{Shapley2016a}
{Shapley}, A.~E., {Steidel}, C.~C., {Strom}, A.~L., {et~al.} 2016, \apjl, 826,
  L24, \dodoi{10.3847/2041-8205/826/2/L24}

\bibitem[{{Sharon} {et~al.}(2020){Sharon}, {Bayliss}, {Dahle}, {Dunham},
  {Florian}, {Gladders}, {Johnson}, {Mahler}, {Paterno-Mahler}, {Rigby},
  {Whitaker}, {Akhshik}, {Koester}, {Murray}, {Remolina Gonz{\'a}lez}, \&
  {Wuyts}}]{Sharon2020a}
{Sharon}, K., {Bayliss}, M.~B., {Dahle}, H., {et~al.} 2020, \apjs, 247, 12,
  \dodoi{10.3847/1538-4365/ab5f13}

\bibitem[{{Shivaei} {et~al.}(2018){Shivaei}, {Reddy}, {Siana}, {Shapley},
  {Kriek}, {Mobasher}, {Freeman}, {Sanders}, {Coil}, {Price}, {Fetherolf},
  {Azadi}, {Leung}, \& {Zick}}]{Shivaei2018a}
{Shivaei}, I., {Reddy}, N.~A., {Siana}, B., {et~al.} 2018, \apj, 855, 42,
  \dodoi{10.3847/1538-4357/aaad62}

\bibitem[{{Siana} {et~al.}(2007){Siana}, {Teplitz}, {Colbert}, {Ferguson},
  {Dickinson}, {Brown}, {Conselice}, {de Mello}, {Gardner}, {Giavalisco}, \&
  {Menanteau}}]{Siana2007a}
{Siana}, B., {Teplitz}, H.~I., {Colbert}, J., {et~al.} 2007, \apj, 668, 62,
  \dodoi{10.1086/521185}

\bibitem[{{Siana} {et~al.}(2010){Siana}, {Teplitz}, {Ferguson}, {Brown},
  {Giavalisco}, {Dickinson}, {Chary}, {de Mello}, {Conselice}, {Bridge},
  {Gardner}, {Colbert}, \& {Scarlata}}]{Siana2010a}
{Siana}, B., {Teplitz}, H.~I., {Ferguson}, H.~C., {et~al.} 2010, \apj, 723,
  241, \dodoi{10.1088/0004-637X/723/1/241}

\bibitem[{{Simmonds} {et~al.}(2024){Simmonds}, {Tacchella}, {Hainline},
  {Johnson}, {McClymont}, {Robertson}, {Saxena}, {Sun}, {Witten}, {Baker},
  {Bhatawdekar}, {Boyett}, {Bunker}, {Charlot}, {Curtis-Lake}, {Egami},
  {Eisenstein}, {Hausen}, {Maiolino}, {Maseda}, {Scholtz}, {Williams},
  {Willott}, \& {Witstok}}]{Simmonds2024a}
{Simmonds}, C., {Tacchella}, S., {Hainline}, K., {et~al.} 2024, \mnras, 527,
  6139, \dodoi{10.1093/mnras/stad3605}

\bibitem[{{Steidel} {et~al.}(2018){Steidel}, {Bogosavljevi{\'c}}, {Shapley},
  {Reddy}, {Rudie}, {Pettini}, {Trainor}, \& {Strom}}]{Steidel2018a}
{Steidel}, C.~C., {Bogosavljevi{\'c}}, M., {Shapley}, A.~E., {et~al.} 2018,
  \apj, 869, 123, \dodoi{10.3847/1538-4357/aaed28}

\bibitem[{{Steidel} {et~al.}(2001){Steidel}, {Pettini}, \&
  {Adelberger}}]{Steidel2001a}
{Steidel}, C.~C., {Pettini}, M., \& {Adelberger}, K.~L. 2001, \apj, 546, 665,
  \dodoi{10.1086/318323}

\bibitem[{{The Astropy Collaboration} {et~al.}(2018){The Astropy
  Collaboration}, {Price-Whelan}, {Sip{\H o}cz}, {G{\"u}nther}, {Lim},
  {Crawford}, {Conseil}, {Shupe}, {Craig}, {Dencheva}, {Ginsburg},
  {VanderPlas}, {Bradley}, {P{\'e}rez-Su{\'a}rez}, {de Val-Borro}, {Paper
  Contributors}, {Aldcroft}, {Cruz}, {Robitaille}, {Tollerud}, {Coordination
  Committee}, {Ardelean}, {Babej}, {Bach}, {Bachetti}, {Bakanov}, {Bamford},
  {Barentsen}, {Barmby}, {Baumbach}, {Berry}, {Biscani}, {Boquien}, {Bostroem},
  {Bouma}, {Brammer}, {Bray}, {Breytenbach}, {Buddelmeijer}, {Burke},
  {Calderone}, {Cano Rodr{\'{i}}guez}, {Cara}, {Cardoso}, {Cheedella}, {Copin},
  {Corrales}, {Crichton}, {D'Avella}, {Deil}, {Depagne}, {Dietrich}, {Donath},
  {Droettboom}, {Earl}, {Erben}, {Fabbro}, {Ferreira}, {Finethy}, {Fox},
  {Garrison}, {Gibbons}, {Goldstein}, {Gommers}, {Greco}, {Greenfield},
  {Groener}, {Grollier}, {Hagen}, {Hirst}, {Homeier}, {Horton}, {Hosseinzadeh},
  {Hu}, {Hunkeler}, {Ivezi{\'c}}, {Jain}, {Jenness}, {Kanarek}, {Kendrew},
  {Kern}, {Kerzendorf}, {Khvalko}, {King}, {Kirkby}, {Kulkarni}, {Kumar},
  {Lee}, {Lenz}, {Littlefair}, {Ma}, {Macleod}, {Mastropietro}, {McCully},
  {Montagnac}, {Morris}, {Mueller}, {Mumford}, {Muna}, {Murphy}, {Nelson},
  {Nguyen}, {Ninan}, {N{\"o}the}, {Ogaz}, {Oh}, {Parejko}, {Parley}, {Pascual},
  {Patil}, {Patil}, {Plunkett}, {Prochaska}, {Rastogi}, {Reddy Janga},
  {Sabater}, {Sakurikar}, {Seifert}, {Sherbert}, {Sherwood-Taylor}, {Shih},
  {Sick}, {Silbiger}, {Singanamalla}, {Singer}, {Sladen}, {Sooley},
  {Sornarajah}, {Streicher}, {Teuben}, {Thomas}, {Tremblay}, {Turner},
  {Terr{\'o}n}, {van Kerkwijk}, {de la Vega}, {Watkins}, {Weaver}, {Whitmore},
  {Woillez}, {Zabalza}, \& {Contributors}}]{Astropy2018a}
{The Astropy Collaboration}, {Price-Whelan}, A.~M., {Sip{\H o}cz}, B.~M.,
  {et~al.} 2018, \aj, 156, 123, \dodoi{10.3847/1538-3881/aabc4f}

\bibitem[{{Thuan} \& {Izotov}(1997)}]{Thuan1997a}
{Thuan}, T.~X., \& {Izotov}, Y.~I. 1997, \apj, 489, 623, \dodoi{10.1086/304826}

\bibitem[{{Vanzella} {et~al.}(2012){Vanzella}, {Guo}, {Giavalisco}, {Grazian},
  {Castellano}, {Cristiani}, {Dickinson}, {Fontana}, {Nonino}, {Giallongo},
  {Pentericci}, {Galametz}, {Faber}, {Ferguson}, {Grogin}, {Koekemoer},
  {Newman}, \& {Siana}}]{Vanzella2012a}
{Vanzella}, E., {Guo}, Y., {Giavalisco}, M., {et~al.} 2012, \apj, 751, 70,
  \dodoi{10.1088/0004-637X/751/1/70}

\bibitem[{{Vanzella} {et~al.}(2016){Vanzella}, {de Barros}, {Vasei}, {Alavi},
  {Giavalisco}, {Siana}, {Grazian}, {Hasinger}, {Suh}, {Cappelluti}, {Vito},
  {Amorin}, {Balestra}, {Brusa}, {Calura}, {Castellano}, {Comastri}, {Fontana},
  {Gilli}, {Mignoli}, {Pentericci}, {Vignali}, \& {Zamorani}}]{Vanzella2016a}
{Vanzella}, E., {de Barros}, S., {Vasei}, K., {et~al.} 2016, \apj, 825, 41,
  \dodoi{10.3847/0004-637X/825/1/41}

\bibitem[{{Vanzella} {et~al.}(2018){Vanzella}, {Nonino}, {Cupani},
  {Castellano}, {Sani}, {Mignoli}, {Calura}, {Meneghetti}, {Gilli}, {Comastri},
  {Mercurio}, {Caminha}, {Caputi}, {Rosati}, {Grillo}, {Cristiani}, {Balestra},
  {Fontana}, \& {Giavalisco}}]{Vanzella2018a}
{Vanzella}, E., {Nonino}, M., {Cupani}, G., {et~al.} 2018, \mnras, 476, L15,
  \dodoi{10.1093/mnrasl/sly023}

\bibitem[{{Vanzella} {et~al.}(2020){Vanzella}, {Meneghetti}, {Pastorello},
  {Calura}, {Sani}, {Cupani}, {Caminha}, {Castellano}, {Rosati}, {D'Odorico},
  {Cristiani}, {Grillo}, {Mercurio}, {Nonino}, {Brammer}, \&
  {Hartman}}]{Vanzella2020a}
{Vanzella}, E., {Meneghetti}, M., {Pastorello}, A., {et~al.} 2020, \mnras, 499,
  L67, \dodoi{10.1093/mnrasl/slaa163}

\bibitem[{{Vasei} {et~al.}(2016){Vasei}, {Siana}, {Shapley}, {Quider}, {Alavi},
  {Rafelski}, {Steidel}, {Pettini}, \& {Lewis}}]{Vasei2016a}
{Vasei}, K., {Siana}, B., {Shapley}, A.~E., {et~al.} 2016, \apj, 831, 38,
  \dodoi{10.3847/0004-637X/831/1/38}

\bibitem[{{Verhamme} {et~al.}(2015){Verhamme}, {Orlitov{\'a}}, {Schaerer}, \&
  {Hayes}}]{Verhamme2015a}
{Verhamme}, A., {Orlitov{\'a}}, I., {Schaerer}, D., \& {Hayes}, M. 2015, \aap,
  578, A7, \dodoi{10.1051/0004-6361/201423978}

\bibitem[{Virtanen {et~al.}(2020)Virtanen, Gommers, Oliphant, Haberland, Reddy,
  Cournapeau, Burovski, Peterson, Weckesser, Bright, {van der Walt}, Brett,
  Wilson, Millman, Mayorov, Nelson, Jones, Kern, Larson, Carey, Polat, Feng,
  Moore, {VanderPlas}, Laxalde, Perktold, Cimrman, Henriksen, Quintero, Harris,
  Archibald, Ribeiro, Pedregosa, {van Mulbregt}, \& {SciPy 1.0
  Contributors}}]{Virtanens2020a}
Virtanen, P., Gommers, R., Oliphant, T.~E., {et~al.} 2020, Nature Methods, 17,
  261, \dodoi{10.1038/s41592-019-0686-2}

\bibitem[{{Wang} {et~al.}(2019){Wang}, {Heckman}, {Leitherer}, {Alexandroff},
  {Borthakur}, \& {Overzier}}]{Wang2019a}
{Wang}, B., {Heckman}, T.~M., {Leitherer}, C., {et~al.} 2019, \apj, 885, 57,
  \dodoi{10.3847/1538-4357/ab418f}

\bibitem[{{Wise} {et~al.}(2014){Wise}, {Demchenko}, {Halicek}, {Norman},
  {Turk}, {Abel}, \& {Smith}}]{Wise2014a}
{Wise}, J.~H., {Demchenko}, V.~G., {Halicek}, M.~T., {et~al.} 2014, \mnras,
  442, 2560, \dodoi{10.1093/mnras/stu979}

\bibitem[{{Witten} {et~al.}(2023){Witten}, {Laporte}, {Martin-Alvarez},
  {Sijacki}, {Yuan}, {Haehnelt}, {Baker}, {Dunlop}, {Ellis}, {Grogin},
  {Illingworth}, {Katz}, {Koekemoer}, {Magee}, {Maiolino}, {McClymont},
  {P{\'e}rez-Gonz{\'a}lez}, {Puskas}, {Roberts-Borsani}, {Santini}, \&
  {Simmonds}}]{Witten2023a}
{Witten}, C., {Laporte}, N., {Martin-Alvarez}, S., {et~al.} 2023, arXiv
  e-prints, arXiv:2303.16225, \dodoi{10.48550/arXiv.2303.16225}

\bibitem[{{Xu} {et~al.}(2016){Xu}, {Wise}, {Norman}, {Ahn}, \&
  {O'Shea}}]{Xu2016a}
{Xu}, H., {Wise}, J.~H., {Norman}, M.~L., {Ahn}, K., \& {O'Shea}, B.~W. 2016,
  \apj, 833, 84, \dodoi{10.3847/1538-4357/833/1/84}

\bibitem[{{Xu} {et~al.}(2022){Xu}, {Henry}, {Heckman}, {Chisholm}, {Worseck},
  {Gronke}, {Jaskot}, {McCandliss}, {Flury}, {Giavalisco}, {Ji}, {Amor{\'\i}n},
  {Berg}, {Borthakur}, {Bouche}, {Carr}, {Erb}, {Ferguson}, {Garel}, {Hayes},
  {Makan}, {Marques-Chaves}, {Rutkowski}, {{\"O}stlin}, {Rafelski},
  {Saldana-Lopez}, {Scarlata}, {Schaerer}, {Trebitsch}, {Tremonti}, {Verhamme},
  \& {Wang}}]{Xu2022a}
{Xu}, X., {Henry}, A., {Heckman}, T., {et~al.} 2022, \apj, 933, 202,
  \dodoi{10.3847/1538-4357/ac7225}

\bibitem[{{Xu} {et~al.}(2023){Xu}, {Henry}, {Heckman}, {Chisholm},
  {Marques-Chaves}, {Leclercq}, {Berg}, {Jaskot}, {Schaerer}, {Worseck},
  {Amor{\'\i}n}, {Atek}, {Hayes}, {Ji}, {{\"O}stlin}, {Saldana-Lopez}, \&
  {Thuan}}]{Xu2023a}
---. 2023, \apj, 943, 94, \dodoi{10.3847/1538-4357/aca89a}

\bibitem[{{Zhang} \& {Bloom}(2020)}]{Zhang2020a}
{Zhang}, K., \& {Bloom}, J.~S. 2020, \apj, 889, 24,
  \dodoi{10.3847/1538-4357/ab3fa6}

\bibitem[{{Zitrin} {et~al.}(2012){Zitrin}, {Rosati}, {Nonino}, {Grillo},
  {Postman}, {Coe}, {Seitz}, {Eichner}, {Broadhurst}, {Jouvel}, {Balestra},
  {Mercurio}, {Scodeggio}, {Ben{\'\i}tez}, {Bradley}, {Ford}, {Host},
  {Jimenez-Teja}, {Koekemoer}, {Zheng}, {Bartelmann}, {Bouwens}, {Czoske},
  {Donahue}, {Graur}, {Graves}, {Infante}, {Jha}, {Kelson}, {Lahav}, {Lazkoz},
  {Lemze}, {Lombardi}, {Maoz}, {McCully}, {Medezinski}, {Melchior},
  {Meneghetti}, {Merten}, {Molino}, {Moustakas}, {Ogaz}, {Patel}, {Regoes},
  {Riess}, {Rodney}, {Umetsu}, \& {Van der Wel}}]{Zitrin2012a}
{Zitrin}, A., {Rosati}, P., {Nonino}, M., {et~al.} 2012, \apj, 749, 97,
  \dodoi{10.1088/0004-637X/749/2/97}

\bibitem[{{Zitrin} {et~al.}(2015){Zitrin}, {Labb{\'e}}, {Belli}, {Bouwens},
  {Ellis}, {Roberts-Borsani}, {Stark}, {Oesch}, \& {Smit}}]{Zitrin2015a}
{Zitrin}, A., {Labb{\'e}}, I., {Belli}, S., {et~al.} 2015, \apjl, 810, L12,
  \dodoi{10.1088/2041-8205/810/1/L12}

\end{thebibliography}
\end{document}